%% file: 0main.tex
\documentclass{IEEEtran}
\usepackage{amsmath,amsfonts}
\usepackage{algorithmic}
\usepackage{algorithm}
\usepackage{array}
\usepackage[caption=false,font=normalsize,labelfont=sf,textfont=sf]{subfig}

\usepackage{textcomp}
\usepackage{stfloats}
\usepackage{url}
\usepackage{verbatim}
\usepackage{graphicx}
\usepackage{cite}

\usepackage{microtype}
\usepackage{booktabs}
\usepackage{hyperref}
\usepackage{amsmath}
\usepackage{amssymb}
\usepackage{mathtools}
\usepackage[capitalize,noabbrev]{cleveref}
\usepackage[utf8]{inputenc}
\usepackage[T1]{fontenc}
\usepackage{nicefrac}
\usepackage{xcolor}
\usepackage{bbm}
\usepackage{enumitem}
\usepackage{makecell}
\usepackage{cuted}
\usepackage{cancel}
\usepackage{custom}
\usepackage{dsfont}

\usepackage[commandnameprefix=always,final]{changes} 

\input{math_commands}

\begin{document}

\title{Neural Distributed Source Coding}

\author{Jay Whang*, Alliot Nagle*, Anish Acharya, Hyeji Kim, \IEEEmembership{Member, IEEE}, Alexandros G. Dimakis, \IEEEmembership{Fellow, IEEE}
\thanks{* denotes equal authorship.}
\thanks{Submitted for review on 12/15/2023. This work was supported by ARO Award W911NF2310062, ONR Award N00014-21-1-2379, NSF Award CNS-2008824, 6G$@$UT center within the Wireless Networking and Communications Group (WNCG), the iMAGiNE Consortium at the University of Texas at Austin, NSF Grants AF 1901292, CNS 2148141, Tripods CCF 1934932, IFML CCF 2019844 and research gifts by Western Digital, Amazon, WNCG IAP, UT Austin Machine Learning Lab (MLL), Cisco and the Stanly P. Finch Centennial Professorship in Engineering.}
\thanks{Jay Whang is with Google DeepMind. Alliot Nagle, Anish Acharya, Hyeji Kim, and Alexandros G. Dimakis are with the ECE department at UT Austin (email: \{jaywhang,acnagle,anishacharya\}@utexas.edu, \{hyeji.kim,dimakis\}@austin.utexas.edu).}
}

\maketitle

\begin{abstract}
We consider the Distributed Source Coding (DSC) problem concerning the task of encoding an input in the absence of correlated side information that is only available to the decoder. Remarkably, Slepian and Wolf showed in 1973 that an encoder without access to the side information can asymptotically achieve the same compression rate as when the side information is available to it. This seminal result was later extended to lossy compression of distributed sources by Wyner, Ziv, Berger, and Tung. 
While there is vast prior work on this topic, practical DSC has been limited to synthetic datasets and specific correlation structures. Here we present a framework for lossy DSC that is agnostic to the correlation structure and can scale to high dimensions. Rather than relying on hand-crafted source modeling, our method utilizes a conditional Vector-Quantized Variational auto-encoder (VQ-VAE) to learn the distributed encoder and decoder. We evaluate our method on multiple datasets and show that our method can handle complex correlations and achieves state-of-the-art PSNR. Our code is made available at \href{https://github.com/acnagle/neural-dsc}{\texttt{https://github.com/acnagle/neural-dsc}}.
\end{abstract}


\input{1intro}
\input{2method}
\input{3experiments}
\input{4discussion}
\input{5conclusion}

\bibliography{bibs/batch_sgd,bibs/distributed_compression,bibs/neural_compression,bibs/sgd_compress,bibs/nit-reference}
\bibliographystyle{IEEEtran}

\newpage
\input{6appendix}

\end{document}

%% file: math_commands.tex
\usepackage{amsmath,amsthm,amssymb,amsfonts}
\usepackage{algorithm}
\usepackage{algorithmic}

\usepackage{comment}
\usepackage{bm}
\usepackage{pifont}

\theoremstyle{plain}

\newtheorem{proposition}{Proposition}

\newtheorem{remark}{Remark}[]

\def\1{\mathbf{1}}

\def\vg{{\mathbf{g}}}

\DeclareMathAlphabet{\mathsfit}{\encodingdefault}{\sfdefault}{m}{sl}
\SetMathAlphabet{\mathsfit}{bold}{\encodingdefault}{\sfdefault}{bx}{n}

\def\gX{{\mathcal{X}}}

\def\sR{{\mathbb{R}}}

\newcommand{\E}{\mathbb{E}}

\newcommand{\KL}{D_{\mathrm{KL}}}

\DeclareMathOperator*{\argmin}{arg\,min}

%% file: 1intro.tex
\section{Introduction}

Data compression plays an essential role in modern computer systems. From multimedia codecs running on consumer devices to cloud backups in large data centers, compression is a necessary component in any system that deals with high-volume or high-velocity sources. 
Applications such as multi-camera surveillance systems, IoT sensing, 3D scene capture and stereo imaging create distributed data streams with very large volume that are highly correlated.  
The central question is how each encoder terminal can efficiently compress its respective data so the decoder can recover the distributed data sources reliably.

This problem has been widely studied in the context of {\em distributed source coding} in information theory. For lossless distributed compression of discrete sources, Slepian and Wolf showed in 1973 that, surprisingly, the sum rate can be reduced to the joint entropy of the sources~\cite{Slepian-Wolf1973}, i.e., the limit on the distributed lossless compression is equivalent to that on centralized compression, where the sources are jointly encoded. Cover provided proof of this surprising result using random binning techniques and extended this result to any pair of stationary ergodic sources~\cite{Cover1975a}.  
For lossy distributed compression, Berger~\cite{Berger1978} and Tung~\cite{Tung1978} extended the Slepian-Wolf scheme to a compress-bin scheme, where typically the encoders perform joint encoding and binding, and the decoder performs joint typically decoding and symbol-by-symbol reconstruction. The Berger-Tung inner bound is not optimal in general~\cite{Wagner--Kelly--Altug2011}; 
however, it is shown to be optimal for several cases, e.g., two correlated Gaussian distributions with mean square error distortion~\cite{Wagner--Tavildar--Viswanath2008}, two correlated binary sources with log loss~\cite{courtade_tsachy_logloss}, the Gaussian CEO problems for two~\cite{Oohama1998,Berger--Zhang--Viswanathan1996,Prabhakaran--Tse--Ramchandran2004} or more~\cite{Prabhakaran--Tse--Ramchandran2004,Oohama2005} sources.

We are interested in a special case of distributed source coding (DSC) -- specifically the problem of \textit{compression with decoder side information}.
As depicted in Figure~\ref{fig:dist-vs-joint} (left), in this setting there are two correlated sources (input $\orig$ and side information $\sinfo$) that 
are physically separated. Both must be compressed and sent to a common decoder, but we assume that the side information $\sinfo$ is compressed in isolation and communicated to the decoder and now one can expect that the original source $\orig$ can be compressed at a higher-rate since $\sinfo$ is known to the decoder.  
The core challenge is how to compress the original source $\orig$ when the correlated side information $\sinfo$ is available only at the decoder as shown in \cref{fig:dist-vs-joint} (left). 
If side information is available to both the encoder and decoder as in \cref{fig:dist-vs-joint} (right), it is well known that the side information can be utilized to improve the compression rate of $\orig$ (Chapter 11 in \cite{ElGamal_Kim_2011}).

Surprisingly, an encoder that has no access to the correlated side information can asymptotically achieve the same compression rate, for lossless compression, as when side information is available at both the encoder and the decoder, which is $H(\orig|\sinfo)$~\cite{Slepian-Wolf1973}. In other words, \textit{distributed} compression is asymptotically as efficient as \textit{joint} compression. 
%
 This is a remarkable result in classical information theory that defies intuition. For lossy compression, the situation is more nuanced. Wyner and Ziv extended the Slepian-Wolf scheme to include quantization, followed by random binning, for lossy compression~\cite{Wyner-Ziv1976}; this extension is a specific case of the Berger-Tung inner bound for distributed source coding~\cite{Berger1978,Tung1978}.  
 Whether the rate-distortion trade-off for the joint encoder is equivalent to that for the distributed encoder, however, depends on the distribution of the source and side information and the distortion metrics. For correlated binary sources with Hamming distortion, providing side information to both the encoder and decoder can reduce the compression rate. However, it does not offer any rate gain for correlated Gaussian sources with mean squared error distortion~\cite{el2011network}.

\input{figures/fig_joint_vs_dist}

\chreplaced{In}{More recently, in} 1999, Pradhan and Ramchandran \chreplaced{advanced}{advance} the insight from the Slepian-Wolf result~\cite{Slepian-Wolf1973} and \chreplaced{introduced}{introduce} a constructive scheme for the distributed coding of i.i.d. correlated sources, such as Bernoulli \chreplaced{sources}{ones}, denoted by \chreplaced{Distributed}{DIstributed} Coding Using Syndromes (DISCUS)~\cite{DISCUS1999}. The key underlying concept, as indicated by its name, is to utilize channel coding and syndrome decoding techniques, which efficiently replace random binning and joint typicality decoding.
This idea can be readily extended to \textit{lossy} distributed compression by first quantizing the source, then applying Slepian-Wolf coding as was proposed by \cite{Wyner-Ziv1976}. Several constructive schemes that combine the DISCUS along with quantization are proposed, e.g., for correlated i.i.d. Gaussian sources, and are shown to be near-optimal as the message length approaches infinity~\cite{DISCUS1999,Yang2003}. 

Despite these theoretical advancements, there are substantial challenges when it comes to designing practical distributed compression schemes for arbitrarily correlated sources.
First, the joint distribution $p(\orig,\sinfo)$ is required to design the encoder and decoder, but modeling high-dimensional joint distributions (e.g. for correlated images) is very challenging, especially without modern deep generative models.  
Second, designing a distributed compression scheme with reasonable run time beyond simple structures \chadded{(e.g., correlated i.i.d. Bernoulli or Gaussian sources)} 
remains an open question.
The significant gap between the theory and practice of DSC has been well-acknowledged in the information theory community. Two decades after the Slepian-Wolf theorem,
~\cite{Verdu1998} writes: ``despite the existence of potential applications, the conceptual importance of Slepian–Wolf coding has not been mirrored in practical data compression.'' Constructive compression schemes have been designed only for very special cases, such as correlated Bernoulli sources~\cite{DISCUS1999}, Gaussian sources~\cite{Yang2003,DISCUS1999}, and stereo images (as elaborated in \cref{sec:related}). 

In this paper, we bridge this gap by leveraging the recent advances in deep generative modeling and show that it is possible to train an encoder and decoder for distributed compression of \textit{arbitrarily} correlated high-dimensional sources. 
Specifically, our approach (denoted \textit{Neural DSC}) parametrizes the encoder-decoder pair as a Vector-Quantized VAE (VQ-VAE) \cite{van2017neural}, so that:
(a) we avoid needing a hand-crafted analytical model for the source distribution; (b) we can \textit{learn} the complicated encoder and decoder mappings via neural networks with efficient inference; and (c) the discrete latent representation of VQ-VAE allows for a further rate improvement through post-hoc training of a latent prior (\cref{sec:our_method}). Our main contributions are as follows:

\begin{itemize}
    \item We introduce Neural DSC, a compression scheme for lossy distributed source coding based on a particular type of VAE (VQ-VAE). The VQ-VAE architecture contains a codebook learned during training in addition to the encoder and decoder, to which the latent representations are quantized.
    \item We justify the use of \chadded{the} VAE by establishing a connection between DSC and a modified evidence lower bound (ELBO) used to train VAEs for our asymmetric encoder-decoder setup. \chadded{We call this modified objective \textsf{dELBO} (\textit{distributed ELBO}) and use it to obtain an objective for training our VQ-VAE.}
    \item We show that Neural DSC performs favorably to existing techniques on stereo camera image compression, achieving state-of-the-art PSNR for rates above 0.1 bit per pixel (bpp); we observe up to a 21\% rate reduction for the same PSNR against the best baseline we compare to. Moreover, our method achieves competitive performance in MS-SSIM \cite{wang2003multiscale}, and we achieve these results with $4.2-7.6\times$ fewer parameters than the baselines.
    \item We further validate that our approach can adapt to complex correlations and other data modalities, such as gradients for distributed learning, and study how effectively the decoder uses side information. \chadded{Our work is novel in this regard since prior work only considers stereo images in their experiments.}
    We also provide a comparison between our Neural DSC and the 
    theoretically optimal compressors on a synthetic dataset. 
\end{itemize}

%% file: figures/fig_joint_vs_dist.tex
\begin{figure}[htbp]
\centering
\includegraphics[width=\linewidth]{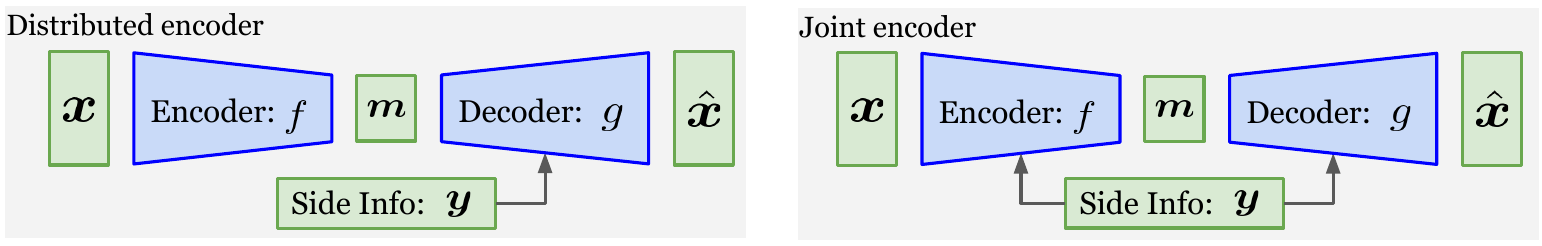}
\caption{A distributed encoder that has no access to the correlated side-information (left) can asymptotically achieve the same compression rate as when side-information is available at both the encoder and the decoder (right).}
\label{fig:dist-vs-joint}
\end{figure}

%% file: 2method.tex
\section{Background}

In this section, we introduce the background necessary to introduce our method. First, we provide a motivating example to provide some intuition on DSC and follow it up by briefly providing some background on ELBO, VAE, and VQ-VAE. Finally, we discuss some related work for learning-based DSC.

\subsection{Intuition behind DSC} \label{sec:motivating_example}

To provide some intuition behind distributed source coding, we describe a simple example that illustrates how side information known \emph{only} to the decoder can be as useful as side information known to \emph{both} the encoder and decoder~\cite{DISCUS1999}.
Let $\orig$ and $\sinfo$ be uniformly random 3-bit sources that differ by at most one bit. 
Clearly, losslessly compressing $\orig$ requires 3 bits. However, if $\sinfo$ is known to both encoder and decoder, then $\orig$ can be transmitted using 2 bits instead. This is because the encoder can send the difference between $\orig$ and $\sinfo$, which is uniform in $\{000,001,010,100\}$. Thus, \textit{joint} compression uses 2 bits.

Now, if the side information $\sinfo$ is available only at the decoder, Slepian-Wolf theorem suggests that the encoder can still transmit $\orig$ using only $2$ bits. 
How could this be possible? The key idea is to group $8$ possible values of $\orig$ into $4$ bins, each containing two bit-strings with maximal Hamming distance: $\mathcal{B}_0 = \{000,111\}$, $\mathcal{B}_1 = \{001,110\}$, $\mathcal{B}_2 = \{010,101\}$, $\mathcal{B}_3 = \{011,100\}$. Then the encoder simply transmits the bin index $\msg \in \set{0,1,2,3}$ for the bin containing $\orig$. The decoder can produce the reconstruction $\reconst$ based on the bin index $\msg$ and $\sinfo$; precisely, 
$\hat{\orig} = \arg\max_{\orig \in \mathcal{B}_{\msg}}  P(\orig|\sinfo)$. Since $\orig$ and $\sinfo$ are off by at most one bit and the Hamming distance between the bit strings in each bin is $3$, the decoder can recover $\orig$ without error given $\sinfo$. In other words, the side information allows the decoder to correctly choose between the two candidates in the bin specified by the encoder.

\subsection{ELBO \& VAE}
A variational auto-encoder (VAE) \cite{kingma2014auto} is a special type of auto-encoder that regularizes the latent space such that the latent variable $\bz$ is a Gaussian random variable. Typically, the training objective of auto-encoders is to minimize the reconstruction loss. However, the training objective of \chreplaced{the VAE}{VAEs} is to recover the true distribution $p(\bx)$ from which the training data is assumed to have been sampled from.

While the decoder $p(\bx\mid\bz)$ allows us to trivially compute $p(\bx,\bz) = p(\bx\mid\bz)p(\bz)$ under the assumption that $\bz$ is a Gaussian random variable, we still cannot directly compute $p(\bx) = \int_{\bz} p(\bx\mid\bz)p(\bz)$ due to the intractable marginalization over $\bz$.  Thus, the VAE training objective instead seeks to maximize the evidence lower bound (ELBO). ELBO is a lower bound on the log-likelihood $p(\bx)$ of the training data that depends on an encoder distribution $q(\bz\mid\bx)$, often also called the variational posterior. The tightness of this lower bound depends on the choice of $q$, with equality achieved when $q$ exactly recovers the true posterior, i.e. $q(\bz\mid\bx) \equiv p(\bz\mid\bx)$.

VAE training is done by \textit{jointly} training both the decoder distribution $p(\bx\mid\bz)$ and the encoder distribution $q(\bz\mid\bx)$ to maximize the ELBO, which consists of a distortion (reconstruction) term and a rate term: 
\begin{equation}
\label{eqn:elbo}
\begin{aligned}
\hspace{-1em}\log p(\bx) &\ge \text{ELBO}(\bx) \\
&= \underbrace{\mathbb{E}_{q(\bz\mid\bx)}\left[\log p(\bx\mid\bz)\right]}_{\text{distortion}} - \underbrace{\kldiv{q(\bz\mid\bx)}{p(\bz)}}_{\text{rate}}
\end{aligned}
\end{equation}

Intuitively, the distortion term represents the reconstruction loss, and the rate term regularizes the latent space such that the posterior $q(\bz \mid \bx)$ is close to the prior $p(\bz)$. Note again that $p(\bz)$ is zero mean isotropic Gaussian. As a special case, when $p(\bx\mid \bz)$ is assumed to be Gaussian, the distortion term becomes the $\ell_2$ reconstruction loss. A derivation of ELBO and a detailed discussion around VAEs can be found in \cite{kingma2019variational}.

VAEs are commonly used for data compression; one such example is bits back coding \cite{townsend2018practical}. Some earlier works \cite{balle2017endtoend,balle2018variational} have demonstrated the the utility of VAEs for image compression, and these works form the backbone of recent works in the DSC setting \cite{NDIC,NDIC-CAM}; we discuss them in Section~\ref{sec:related}.

\subsection{Vector-Quantized VAE}
\label{sec:vqvae}

Vector-Quantized VAE (VQ-VAE) \cite{van2017neural} is a specific type of VAE \cite{kingma2014auto,rezende2014stochastic} with a \textit{discrete} latent variable, even when the input is continuous.  Because the latent code is discrete and has fixed size, VQ-VAEs are a natural fit for lossy compression.  Indeed, many existing works have explored its use in various compression tasks, ranging from music generation to high-resolution image synthesis \cite{garbacea2019low,razavi2019generating,roberts2018hierarchical,ramesh2021zero,yu2022scaling}.

A VQ-VAE consists of three components: an encoder, a decoder, and a codebook.  The main difference between VQ-VAE and a regular VAE is that the output of the encoder is quantized to the nearest vector in the codebook. During training, all three components are jointly optimized.
During inference, \chreplaced{the encoder compresses the input into a lower-dimensional representation of vectors which are quantized to the latent vectors in the codebook. These quantized representations are passed through the decoder, which reconstructs the image.}{the input is represented by the index of the code vector that the encoder's output is quantized to.}
Once a VQ-VAE is trained, it is common to use an autoregressive model, such as PixelCNN \cite{oord2016conditional} or transformers \cite{vaswani2017}, to achieve a lower rate for a fixed distortion value. The autoregressive model learns the joint distribution for the ordered set of latent codebook vectors used to represent each \chreplaced{input to the VQ-VAE}{VQ-VAE input}; this model can be used to lower the compression rate of the VQ-VAE via arithmetic coding \cite{langdon1984introduction}.

\subsection{Related Work}\label{sec:related}

The authors of DSIN \cite{ayzikA2020dsin} propose a method to perform DSC for stereo camera image pairs with a high spatial correlation. Due to the large spatial similarity between the images, one of them can serve as the side information for the other.
The key component of their method is the ``Side Information Finder'', a module that finds similar image patches between the side information and the reconstructed signal produced by a pre-trained autoencoder. Since the two images have many approximately overlapping patches, the reconstruction is further improved by copying over matching image patches from the side information to the reconstruction. 

While this leads to a considerable improvement in compression performance, this method is only applicable when the input and side information have large spatial overlap.  On the other hand, our method is applicable to any correlated sources. We empirically validate this using data sources with substantially more complex correlation (see \cref{fig:sample_stereo_vs_face}, right).

NDIC \cite{NDIC} leverages a (regular) VAE for DSC of image data; the proposed architecture is designed to explicitly model the common information between the input and side information. Intuitively, the goal is to guide the encoder to compress only the portion that is not recoverable from the side information. Once trained, the encoder simply discards the common information and only transmits the residual, hoping the decoder can reconstruct the common information from the side information. NDIC-CAM \cite{NDIC-CAM} further adapts this architecture by encoding and decoding the side information in parallel to the input image decoder network and utilizing cross-attention modules between the two decoder networks. They report that the cross-attention modules allow for better alignment between the latent representations of the side information and the input image, thereby enabling better utilization of the side information over NDIC. 

LDMIC \cite{zhang2023ldmic} presents a DSC framework for \textit{multi-view} images, where $K$ overlapping images (from multi-view cameras, for example) are compressed independently but decompressed jointly. Their key architectural contribution is their joint context transfer module; this module is used in the decoder to reconstruct image $k\in \{1,\ldots, K\}$ based on the compressed representations of the remaining $\{1, \ldots, K\}\setminus \{k\}$ images. \chadded{They consider a setting where the side information is lossily compressed before being presented to the decoder. This is different from our setting, where we (along with the other methods) consider a situation where the side information is \textit{losslessly} presented to the decoder.}

Finally, a recent study has shown that neural distributed compressors learn to bin their source data \cite{ozyilkan2023neural}. In addition to their provided theory, the authors empirically observe the emergence of binning for Gaussian and Laplacian data. Our work differs in that (1) we study the DSC with complex correlations between the encoder input and the decoder side information, and (2) we consider vector input data, whereas this other work considers scalar input data.

\section{Neural Distributed Source Coding}

In this section, we derive our ELBO bound for the DSC setting. We discuss the connection of our distributed ELBO bound to the optimization objective of NDIC-CAM \cite{NDIC-CAM}, the state-of-the-art baseline we compare our work to. We use our distributed ELBO bound to motivate a new loss for the VQ-VAE architecture. \cref{fig:vqvae} provides a high-level overview of our method, showing how side information is incorporated into the VQ-VAE architecture and how arithmetic coding is used to achieve improved compression rates.

\input{figures/fig_vqvae}

\subsection{ELBO and Distributed Source Coding} 
VAEs have been extensively used for neural compression \cite{balle2017endtoend,minnen2018joint,balle2018variational}, as its training objective, ELBO, can be interpreted as a sum of rate and distortion of a lossy compressor as shown in \cref{eqn:elbo}.
For example, when the decoder $p(\bx\mid\bz)$ is Gaussian, the first term turns into the widely-used $\ell_2$ distortion.

However, this interpretation breaks down in our asymmetric encoder-decoder setup where only the decoder has access to the side information $\by$. In this setup, we are are interested in modelling the conditional likelihood $p(\bx\mid\by)$ as opposed to the likelihood $p(\bx)$ for the symmetric (vanilla VAE) case. We can show that the corresponding objective in our setup, which we call \textit{distributed ELBO} (or \textsf{dELBO} for short), is in fact a lower bound to the conditional log-likelihood $\log p(\orig \mid \sinfo)$:

\begin{proposition}
\label{prop:distributed_elbo}
Let $\orig, \sinfo, \bz$ be random variables following the generative process $\sinfo \rightarrow \orig \leftarrow \bz$, i.e., $\bm{z}$ is the latent variable that is independent of $\bm{y}$ and $p(\bm{x},\bm{y},\bm{z}) = p(\bm{x}|\bm{y}, \bm{z})p(\bm{y})p(\bm{z})$. Then for any choice of posterior $q(\bz\mid\bx)$ valid under $p$ (i.e. $\supp q(\bz\mid\bx) \subseteq \supp p(\bz)$ for all $\bx$), we have
\[
\begin{aligned}
\log p(\bx\mid\by) &\ge 
\underbrace{\mathbb{E}_{q}\left[\log p(\bx\mid\by,\bz)\right]}_{\textnormal{distortion with side info}} - 
\underbrace{\mathrm{KL}\left[q(\bz\mid\bx) \parallel p(\bz)\right]}_{\textnormal{rate with side info}} \\
&\triangleq \textnormal{\textsf{dELBO}}(\bx,\by).
\end{aligned}
\]
\end{proposition}

\noindent \textbf{Proof:}
\begin{align*}
\log p(\orig\mid\sinfo) 
&= \E_{\bz\sim q(\cdot\mid\orig)} \brac{\log p(\orig\mid\sinfo) } \\
&= \E_q\brac{\log p(\orig,\sinfo,\bz) - \log p(\bz\mid\orig,\sinfo) - \log p(\sinfo)}  \\
\begin{split}
    &= \E_q\left[(\log p(\orig,\sinfo\mid\bz) + \log p(\bz)) \right. \\
    &\qquad + \left.(\log q(\bz\mid\orig) - \log q(\bz\mid\orig))\right] \\
    &\qquad - \E_q\brac{\log p(\bz\mid\orig,\sinfo) + \log p(\sinfo)}
\end{split} \\
\begin{split}
    &= \E_q\left[
    -(\log q(\bz\mid\orig) - \log p(\bz))\right. \\
    &\qquad + \left.(\log q(\bz\mid\orig) - \log p(\bz\mid\orig,\sinfo))\right] \\
    &\qquad + \E_q\brac{\log p(\orig,\sinfo\mid\bz) - \log p(\sinfo)}
\end{split} \\
\end{align*}
\begin{align*}
\begin{split}
    &= -\kldiv{q(\bz\mid\orig)}{p(\bz)} \\
    &\qquad + \kldiv{q(\bz\mid\orig)}{p(\bz\mid\orig,\sinfo)} \\
    &\qquad + \E_q\brac{\log p(\orig,\sinfo\mid\bz) - \log p(\sinfo)}
\end{split} \\
\begin{split}
&\ge -\kldiv{q(\bz\mid\orig)}{p(\bz)} \\
&\qquad + \E_q\brac{\log p(\orig,\sinfo\mid\bz) - \log p(\sinfo)}
\end{split} \\
\begin{split}
&\stackeq{A} -\kldiv{q(\bz\mid\orig)}{p(\bz)}
+ \E_q\left[\log p(\orig\mid\sinfo,\bz) \right. \\
&\qquad + \left.\cancel{\log p(\sinfo\mid\bz)} - \cancel{\log p(\sinfo)}\right]
\end{split} \\
&= -\kldiv{q(\bz\mid\orig)}{p(\bz)}
+ \E_q\brac{\log p(\orig\mid\sinfo,\bz)} \\
&= \textsf{dELBO}(\bx,\by)
\end{align*}
where ($A$) follows from our assumption that $\sinfo$ and $\bz$ are marginally independent. \chdeleted{Negating both sides of the inequality completes the proof.}
Note that the \textsf{dELBO} rate term, $\kldiv{q(\bz\mid\orig)}{p(\bz)}$, matches the rate term in ELBO. Intuitively, this makes sense since the rate is controlled by the encoder, which does not have access to the side information $\by$. The DSC setting provides the decoder with access to correlated side information $\by$, and that is reflected in the \textsf{dELBO} distortion term.

The importance of this connection between ELBO and distributed source coding is twofold. First, it shows that there is a tractable surrogate objective that minimizes the optimal conditional compression rate $-\log p(\orig\mid\sinfo)$, which is also the asymptotically optimal rate for distributed coders \cite{Slepian-Wolf1973,Wyner-Ziv1976}.
Second, and perhaps more importantly, this scheme also allows us to obtain a practically useful encoder-decoder pair whose rate-distortion we can control as we shall see later.

\label{remark:obj_fcn}
\begin{remark}\textnormal{ 
We point out that \textsf{dELBO} is a special case of the loss objective proposed in Equation 1 of the NDIC-CAM work \cite{NDIC-CAM}. Beyond the \textsf{dELBO} terms, their loss objective includes terms for the rate and distortion of the side information $\by$, controlled by hyperparameter $\alpha$, and the rate of their proposed \textit{common information}, controlled by hyperparameter $\beta$. \chreplaced{However, they empirically show that setting $\alpha = \beta = 0$ yields the best results, which is the case that recovers our \textsf{dELBO} objective }{Note that their objective is \textsf{dELBO} when $\alpha = \beta = 0$, and they empirically show that this case yields the best performance}. Specifically, their loss function is the following:
\begin{equation}
    L = R_x + \lambda D_x + \alpha \left( R_y + \lambda D_y \right) + \beta R_w,
\end{equation}
where $R_x, D_x, R_y, D_y$ are the rate and distortion terms for the input and side-information, respectively, and $R_w$ is the rate of the \textit{common information}, which is a nonlinear transform of the side-information. Our \textsf{dELBO} objective is recovered for $\alpha = \beta = 0$, since $R_x, D_x$ both depend on $\bx$ given $\by$.}

\end{remark}

\subsection{Our Method}
\label{sec:our_method}

\vspace{.2em}\subsubsection{Notation and Setup} We let $\orig$, $\sinfo$, and \chreplaced{$c_{\msg}$}{$\msg$ to} denote the original message, correlated side information, and compressed message, respectively. The \textit{encoder} $\fencode$ maps the given message to a low-dimensional vector $\fencode(\orig)$, which is then quantized to the \chreplaced{nearest codebook}{closest code} vector:
$$c_{\msg} = \argmin_{c \in \mathcal{C}} \norm{c - \fencode(\orig)},$$ where the table of code vectors $\mathcal{C}$ is jointly trained with the encoder $\fencode$ and the decoder $\fdecode$. \chreplaced{Since the decoder also has access to the codebook, only the index $\msg$ of the compressed message is transmitted, which is accomplished in a lossless manner}{We then transmit the index $\msg$ of the quantized vector losslessly}.
The \textit{decoder} $\fdecode$ in turn tries to reconstruct the input signal using the compressed message and side info: $\reconst = \fdecode(c_{\msg}; \sinfo)$.
Note that the encoder only \chreplaced{accepts $\orig$ as input}{receives $\orig$}, while the decoder \chreplaced{accepts both $c_{\msg}$}{receives both $\msg$} and the side information $\sinfo$. 

We refer to the number of bits required to transmit \chreplaced{$c_{\msg}$}{$\msg$} as the \textit{rate}, and the reconstruction performance (measured by $\ell_2$ error or MS-SSIM) as the \textit{distortion}. \chadded{For our image compression experiments, the rate is reported as bits-per-pixel (bpp) as done in standard practice. bpp measures the average number of bits needed to compress a pixel for a given image. Therefore, the total number of bits required to transmit $c_{\msg}$ (as an RGB image), with $n$ pixels and bpp $r$, is $n \times r$.}

\vspace{.2em}\subsubsection{Training Objective}
\label{sec:train_obj}
We train the distributed encoder-decoder pair $(\fencode, \fdecode)$ along with the codebook $\mathcal{C}$ as a conditional VQ-VAE, where only $\fdecode$ is conditioned on the side information $\sinfo$. 
\chadded{In particular, we {\em iterate} the updates of the distributed encoder-decoder pair $(\fencode, \fdecode)$ and the codebook $\mathcal{C}$; we first discuss the loss function for updating  $(\fencode, \fdecode)$ for a given codebook $\mathcal{C}$, followed by the method for updating $\mathcal{C}$ for a given $(\fencode, \fdecode)$. } 

\begin{equation}
\label{eqn:vqvae_loss}
\begin{aligned}
\mathcal{L}(\fencode, \fdecode, \mathcal{C}) &= 
\norm{\fdecode(c_\msg; \sinfo) - \orig}^2 \\
&\quad + \beta \norm{\fencode(\orig) - \texttt{stop\_grad}(c_\msg)}^2,
\end{aligned}
\end{equation}
\chadded{where the \texttt{stop\_grad} operator acts as the identity function in the forward direction but prohibits the propagation of gradients during training.} This loss can be interpreted as a weighted sum of \chadded{a} log-likelihood term under \chadded{a} Gaussian decoder (distortion) and a vector quantization term that tries to bring encoder output close to existing code vectors, which in turn has an impact on the resulting rate. This training objective is derived from \textsf{dELBO} in a similar fashion as the original VQ-VAE objective was derived from ELBO \cite{van2017neural} \chreplaced{, which we detail here. 
Recall that the \textsf{dELBO} objective we derived for a VAE is 
\[
\textsf{dELBO}(\bx,\by) = -\kldiv{q(\bz\mid\orig)}{p(\bz)}
+ \E_q\brac{\log p(\orig\mid\sinfo,\bz)}.
\] 
For the VQ-VAE objective, first, we assume that $p(\bz)$ is uniform. Consequently, the KL divergence $\kldiv{q(\bz\mid\bx)}{p(\bz)}$ is 
therefore constant with respect to the encoder parameters and can be removed from the objective. In particular, when $p(\bz)$ is uniform, then $\kldiv{q(\bz\mid\bx)}{p(\bz)} = \sum_{z\in\text{supp }q(\bz\mid\orig)} q(\bz\mid\orig) \log \frac{q(\bz\mid\orig)}{1 / |\mathcal{C}|} = \log |\mathcal{C}|$ since $q(\bz\mid\bx)$ is deterministic (a one-hot encoded distribution) when $\orig$ is known. This leaves us with the $\E_q\brac{\log p(\orig\mid\sinfo,\bz)}$ distortion term, which, under the standard assumption that $p(\orig \mid \sinfo, \bz)$ is Gaussian, becomes the MSE loss. This corresponds to the first term in our loss function shown in \cref{eqn:vqvae_loss}. In practice, the distortion does not need to be restricted to MSE and can instead be any loss that measures the distortion between two images, such as MS-SSIM. Next, we include the second term in our loss function shown in \cref{eqn:vqvae_loss}, known as the commitment loss, which forces the encoder to learn to use the codebook and ``commit'' to outputting vectors that are close to the codebook vectors.}

\chreplaced{
Note that our loss does not include a term that reflects codebook updates. Instead, to update the codebook $\mathcal{C}$ for a given $(f,g)$, we follow the strategy of the VQ-VAE paper where the codebook is updated using exponential moving averages (EMA). Intuitively, this approach is similar to a K-means update where the cluster centers (the codebook vectors in our case) are updated according to the average of the nearest vectors. Optimally, each codebook vector $c_i$ shall be the average of the $n_i$ closest output vectors from the encoder across all inputs $\orig$. However, this approach is not well-suited for us, as we only have access to mini-batches.  
Instead, we update each codebook vector $c_i$ as the exponential moving averages of the cluster centers of mini-batch outputs; specifically, we update the $i$-th codebook vector to be $c_i^{(t)} = \frac{\alpha_i^{(t)}}{\beta_i^{(t)}}$, where $\alpha_i^{(t)} = \gamma \alpha_i^{(t-1)} + (1-\gamma)\sum_j e_{i, j}^{(t)}$ and $\beta_i^{(t)} = \gamma \beta_i^{(t-1)} + (1-\gamma) n_i^{(t)}$. Here, $\{e_{i, 1}^{(t)}, e_{i, 2}^{(t)}, \ldots, e_{i, n_i^{(t)}}^{(t)}\}$ are the $n_i^{(t)}$ encoder outputs in the current $t$-th mini-batch which are closest to codebook vector $c_i^{(t)}$. $\gamma\in\left[0, 1\right]$ is a hyperparameter; we follow the original VQ-VAE work and set it to $\gamma = 0.99$. At initialization, prior to any training, $\alpha_i^{(0)}$ is randomly initialized and $N_i^{(0)} = 0$.}{Namely, we assume that $p(\bz)$ is uniform and $\kldiv{q(\bz\mid\bx)}{p(\bz)}$ is therefore constant with \chreplaced{respect}{resepect} to the encoder parameters and can be removed from the objective, we add the commitment loss (the second term in Equation \ref{eqn:vqvae_loss}), and \chadded{the} codebook $\mathcal{C}$ is jointly updated as an exponential moving average of code vectors following the original VQ-VAE work.}

\vspace{.2em}\subsubsection{Why VQ-VAE?}
Although much of the existing literature has focused on VAEs as the backbone for neural compression \cite{balle2017endtoend,theis2017lossy,balle2018variational,minnen2018joint,balle2021nonlinear}, we intentionally chose VQ-VAE instead for a few reasons.
First, VQ-VAE offers explicit control over the rate through the latent dimension and codebook size -- unlike VAEs, for which we can only estimate the rate after training the model. VQ-VAE also guarantees an upper bound on the resulting rate regardless of the input, which is often desired in practice (e.g., a communication channel with a strict maximum bandwidth limit). Finally, a rate improvement can be achieved for a trained VQ-VAE by fitting a latent prior model to the VQ-VAE's discrete latent space; we demonstrate such a rate improvement in \cref{fig:kitti_bpp_psnr}. The downside of using a VQ-VAE is that training is less stable than a VAE, but this did not pose much issue during our experiments. \chadded{These stability issues are remedied through the use of the commitment loss term and the EMA codebook updates, which we discuss further in \cref{sec:train_obj}.}

\vspace{.2em}\subsubsection{Use of a Latent Prior}
Our VQ-VAE architecture achieves comparable performance to other methods in the DSC setting. However, the aggregate posterior learned by VQ-VAE is often far from uniform in practice. This allows for a further rate improvement by losslessly compressing the latent codes by fitting a discrete prior over them. Following \cite{van2017neural}, we use an autoregressive prior based on \chadded{the} Transformer \cite{vaswani2017}, which can be readily used with arithmetic coding \cite{langdon1984introduction} to obtain a lossless compressor that closely matches the model's negative log-likelihood.

Specifically, we use the decoder-only Transformer trained with a maximum-likelihood objective on discrete latent codes. 
This differs from the original VQ-VAE work \cite{van2017neural}, which utilized a PixelCNN prior. \chadded{The} Transformer has been shown to achieve better performance compared to PixelCNN variants \cite{chen2018snail,razavi2019generating}. \chreplaced{In \cref{fig:kitti_bpp}, we provide our VQ-VAE results with (solid blue line) and without (dashed blue line) the latent prior}{We provide VQ-VAE results with (solid blue line) and without (dashed blue line) the latent prior, as shown in Figure 3}. When we use the latent prior, we outperform all other PSNR methods, and match the best method on MS-SSIM.

%% file: figures/fig_vqvae.tex
\begin{figure*}[!h]
\centering
\includegraphics[width=1.0\textwidth]{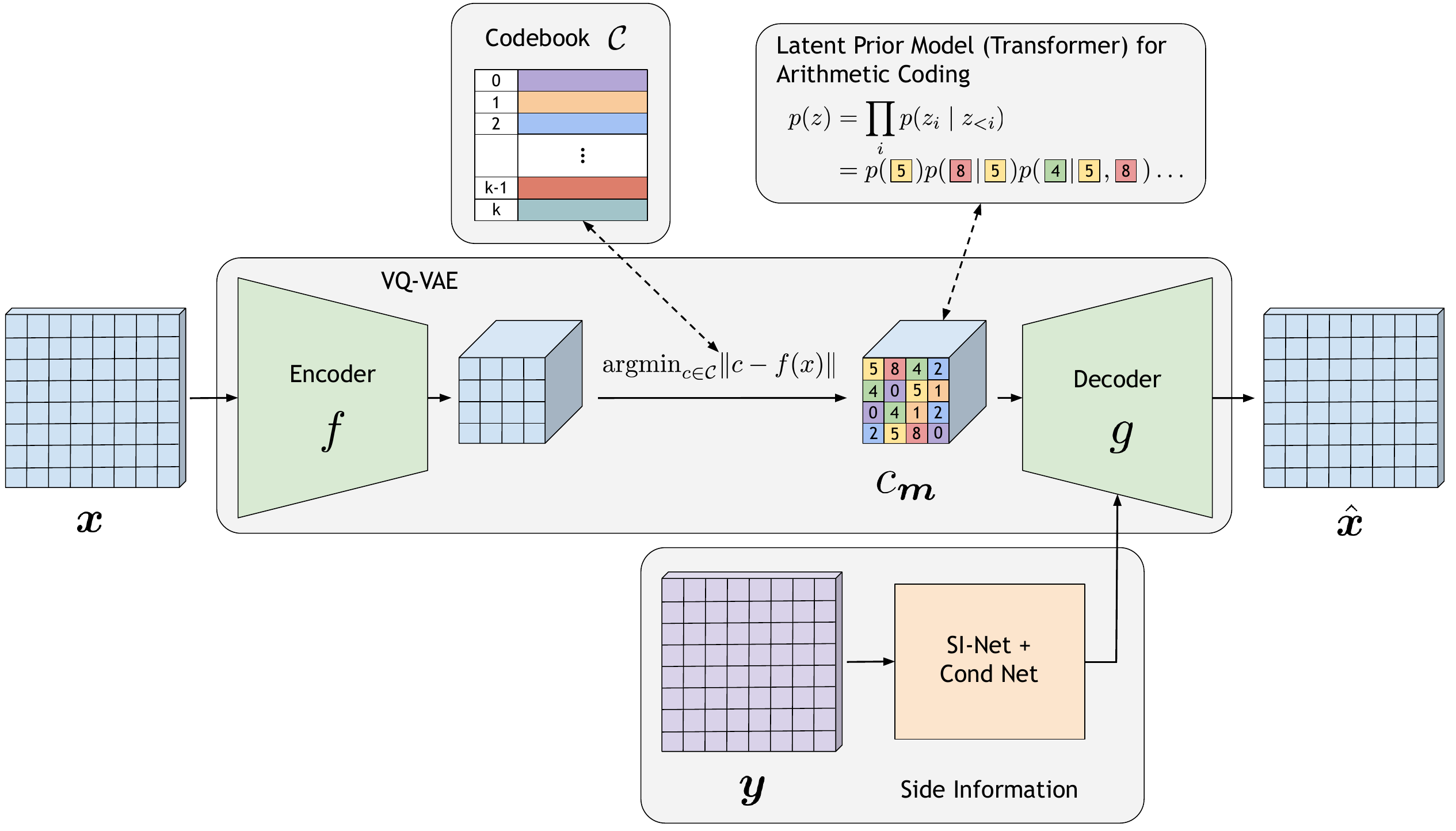}
\caption{We use a modified VQ-VAE architecture with correlated side information at the decoder. We train a prior model over the quantized latents to achieve better compression rates via arithmetic coding. \chadded{The \texttt{SI Net + Cond Net} is a convolutional neural network with residual connections and is similar to the encoder in its architecture. Figures with more details on the exact architecture of the encoder and decoder, and the \texttt{SI Net + Cond Net} are shown in \cref{fig:vqvae_arch_conv} and \cref{fig:si_net_arch_conv}, respectively.}}
\label{fig:vqvae}
\end{figure*}

%% file: 3experiments.tex
\section{Experiments}

We demonstrate the efficacy of our framework through a diverse set of experiments. In our stereo image compression experiments in \cref{sec:stereo_img_comp}, we match or exceed the performance of all baselines. Then in \cref{sec:complex_corr}, we methodically investigate how well VQ-VAEs handle complex correlations between the input message and side information in the distributed, joint, and separate cases. Finally, in \cref{sec:gradients}, we provide a proof of concept experiment to show how our framework can be used in distributed training.

\subsection{Stereo Image Compression}
\label{sec:stereo_img_comp}

\vspace{.2em}\subsubsection{Setup}
We first evaluate our method on stereo image compression.  Following \cite{ayzikA2020dsin}, we construct a dataset consisting of pairs of images obtained from the KITTI Stereo 2012 and 2015 \cite{geiger2012we,menze2015joint}. Each pair of images are taken by two cameras at a slightly different \chreplaced{angles}{angle} and share spatial similarity (see \cref{fig:sample_stereo_vs_face}, left). The goal is to compress one of the images in each pair, treating the other image as side information only the decoder can access. The performance of a compressor is evaluated by its rate-distortion points, where the rate is measured in terms of \chreplaced{bits-per-pixel}{bits/pixel} (bpp) and distortion is measured in terms of Peak Signal-to-Noise Ratio (PSNR) or Structural Similarity (MS-SSIM) \chadded{\cite{wang2003multiscale}}. \chadded{These rate and distortion measures are commonly reported in the image compression literature. More specifically, for an image $I$, it's reconstruction has $\text{PSNR}_I := 10 \log_{10}\left( \frac{\text{MAX}_I}{\text{MSE}_I} \right)$, where $\text{MAX}_I$ is the maximum possible pixel value (which is 255 for an 8-bit representation) and $\text{MSE}_I$ is the Mean-Squared Error (MSE) between each pixel in the reconstruction and $I$. The PSNR metric has the same semantic meaning as the MSE metric, but is expressed on a decibel scale. The MS-SSIM metric is a similarity metric between two images; in our case, we measure the similarity between a ground-truth image and its reconstruction. The exact definition is deferred here due to its complexity compared to PSNR; however, the key idea is that it quantifies the \textit{perceptual} change in \textit{structural information} (where pixels rely on their neighbors to create specific structures) between two images. In contrast, PSNR assesses the \textit{absolute} change between the images. The MS-SSIM metric tends to agree with human perception of distortion better than PSNR \cite{ayzikA2020dsin,ndajah2010ssim}.}

We train a conditional VQ-VAE for each target rate as described in \cref{sec:our_method}. We then evaluate its distortion averaged over the test set and plot the rate-distortion point. The exact details of how the dataset is constructed, as well as the model hyperparameters, are included in the Appendix \chadded{(\cref{app:vq_vae_arch})}.

\vspace{.2em}\subsubsection{Results}
We compare our methods to existing distributed image compression methods DSIN \cite{ayzikA2020dsin}, NDIC \cite{NDIC}, and NDIC-CAM \cite{NDIC-CAM}. While the authors of NDIC-CAM did not report PSNR in their paper, we used their code to train their model with an MSE objective for a fair comparison. We report the $\alpha, \beta, \lambda$ hyperparameters we used to train their \chreplaced{models}{method} in the Appendix \chadded{(\cref{app:ndic_cam})}.

In \cref{fig:kitti_bpp_ssim}, we see that our method remains competitive compared to NDIC-CAM, which currently achieves the best results for MS-SSIM. However, in \cref{fig:kitti_bpp_psnr}, our method outperforms all baselines in PSNR for rates above 0.1 bpp and is competitive with NDIC-CAM for rates below 0.1 bpp. These results suggest that our approach is amenable to \chreplaced{training with different distortion functions}{different objective functions} while maintaining state-of-the-art reconstruction performance. \chadded{In \cref{fig:kitti_bpp_ldmic} in the Appendix, we show \cref{fig:kitti_bpp} but with LDMIC \cite{zhang2023ldmic} included. The setting of LDMIC is different than the remaining methods (including ours) because $\sinfo$ is presented to the decoder after being lossily compressed, whereas the other methods have access to $\sinfo$ directly. Therefore, we defer the result to the Appendix. 
}

\input{figures/fig_kitty_plot_and_samples}

In addition, our model is significantly more parameter-efficient compared to NDIC and NDIC-CAM. \cref{table:param_efficiency} shows the parameter count for the models used to generate the points in \cref{fig:kitti_bpp_psnr}.  As shown, our models are smaller by a factor of up to approximately $8 \times$.

\input{figures/table_param_efficiency}

\subsection{Handling Complex Correlation}
\label{sec:complex_corr}

\vspace{.2em}\subsubsection{Setup}
To further investigate how well our method can handle complex correlations between the input and side information, we evaluate our method on a challenging distributed compression setup. First, we create a dataset of correlated images from $256 \times 256$ CelebA-HQ dataset \cite{liu2015deep} containing images of celebrity faces. Each image is vertically split into top and bottom halves, where the top half is used as the input, and the bottom half is used as side information (see \cref{fig:sample_stereo_vs_face}, right). Following \cite{kingma2018glow}, we use $27000/3000$ split between train/test data.  

While there is clearly some correlation between the top and bottom halves of an image of a human face, modeling this correlation (e.g., the conditional distribution over the top half of a face given the bottom half) is highly nontrivial. This experiment is thus designed to show our model's ability to leverage this correlation to improve compression.

\vspace{.2em}\subsubsection{Baseline VQ-VAEs}
To analyze the gains from distributed compression, we train three different variants of VQ-VAEs: distributed (our method), joint, and separate.  In the \textit{joint} model, both the encoder and decoder have access to the side information\chadded{, as depicted in \cref{fig:dist-vs-joint}}. This serves as a proxy to the intractable theoretical rate-distortion bound established by \cite{Wyner-Ziv1976} for lossy distributed compression.  We expect this to be the upper limit on the performance of our method.  The \textit{separate} model is positioned at the other extreme, where neither the encoder nor the decoder uses the side information.  This serves as the lower limit on the performance of our method.
To ensure a fair comparison among these variants, they have identical architecture and number of parameters for the autoencoder portion and only vary in the way they handle the side information. Architectural details and network hyperparameters are provided in the Appendix \chadded{(\cref{app:vq_vae_arch})}.

\vspace{.2em}\subsubsection{Results}
In \cref{fig:celeba_rate_distortion}, we see that the distributed VQ-VAE achieves nearly identical performance as the joint VQ-VAE, further proving the viability of our method as a distributed coding scheme.
 
\input{figures/fig_celeba_rate_distortion}

\vspace{.2em}\subsubsection{Effect of Side Information}

It is possible that the distributed VQ-VAE learns to ignore the side information, effectively collapsing to a separate VQ-VAE.  We investigate whether the distributed decoder actually uses the side information by intentionally providing incorrect input.

We show in \cref{fig:effect_sideinfo} that the side information plays a significant role in the quality of reconstruction. For example, providing side information with a different face leads to the reconstruction having a matching skin tone that is different from the original. \chdeleted{As expected, side information has no effect for the separate encoder}.

\input{figures/fig_effect_sideinfo}

\subsection{Communication Constrained Distributed Optimization} \label{sec:gradients}

\vspace{.2em}\subsubsection{Background} 
Here, we consider an interesting proof of concept that applies our method to distributed training of a neural network $\net$ with parameters $\params \in \sR^d$.
With the increasing size of deep learning models, 
an important bottleneck in distributed training is the repeated communication of gradients between workers and the parameter server. \chreplaced{Many gradient compression techniques have been developed to alleviate this issue, and here we demonstrate that our approach can be used and it performs well.}{To alleviate this issue, many gradient compression techniques have been developed}.

A key observation we make is that the gradients coming from different workers are correlated. This suggests that individual workers may be able to compress their gradients via distributed source coding without having to communicate with each other. We compare this approach to a representative subset of baselines, which we describe in \cref{sec:sgd_baselines}. While our approach is not ready for practical deployment, we observe a vastly improved training performance, suggesting a promising direction for future research.

\vspace{.2em}\subsubsection{Experimental Setup}

We focus on a simple setup where two workers $j=1,2$ are assigned a subset $\gX_j \in \gX$ of the training data. The workers then compute gradients locally using SGD-like iterations and communicate the gradients back to a central parameter server.
At each iteration $t$, the correlation between the client gradients $\vg_t^1$ and $\vg_t^2$ can be exploited to improve compression performance further.
Concretely, we treat  $\vg_t^1$ as side information, train a distributed encoder for $\vg_t^2$, and show that the cost of communicating $\vg_t^2$ can be substantially reduced.

The neural network $\net$ being trained using compressed gradients is a small convolutional network for MNIST digit classification \cite{lecun2010mnist}.
We partition the MNIST training data $\gX$ into two equal subsets: $\gX_{pre}$ and $\gX_{train}$. $\gX_{pre}$ is used to train the VQ-VAE encoder and $\gX_{train}$ is used to train $\net$ using the trained encoder as gradient compressor. 
We measure the performance of our method using two metrics: (a) \textbf{rate-distortion} 
of the distributed compressor, and (b) \textbf{classification accuracy} of the model $\net$ trained using \chreplaced{gradients compressed by our VQ-VAE}{compressed gradients}.

\vspace{.2em}\subsubsection{Generating training data for \chreplaced{VQ-VAE gradient compressors}{encoder}}
We train $\net$ for $T$ steps using the Adam optimizer
across two nodes over $\gX_{pre}$.  This generates a sequence of $T$ gradient pairs $\set{\vg_t^1, \vg_t^2}_{t=1}^T$, which can be used to train the VQ-VAE \chreplaced{gradient compressors}{encoders}.

However, applying our method to this setting naively leads to a suboptimal VQ-VAE as these gradient pairs are highly correlated across training steps. Noticing that $\vg_t^1$ and $\vg_t^2$ are conditionally independent given the initial model weights and the time step $t$, we train $\net$ for multiple runs with different initialization for $\params$, while randomly sampling a subset of gradients from each run. Thus, we generate a dataset of tuples $(t, \vg_t^1, \vg_t^2)$ sampled from multiple independent runs.  We also update the architecture of the VQ-VAE so that both the encoder and decoder are conditioned on $t$.  This way, the gradients become i.i.d. samples over random runs and time steps.


\vspace{.2em}\subsubsection{Pre-training VQ-VAE \chreplaced{ gradient compressors}{encoders}}
We first compare the performance of three different VQ-VAE variants \chadded{(distributed, joint, and separate, as depicted in \cref{fig:dist-vs-joint})} over the course of a single training run of $\net$. As shown in \cref{fig:grad_vqvae_mse}, we observe a substantial improvement in $\ell_2$ distortion for the distributed VQ-VAE, compared to the separate model. As training progresses, the gap between the joint and distributed encoders narrows, suggesting that a sufficiently large distributed VQ-VAE \chreplaced{compressor}{encoder} can nearly match the performance of its joint counterpart \chadded{in terms of the gradient reconstruction quality (MSE).}

\input{figures/fig_all_grad_plots}

\vspace{.2em}\subsubsection{Distributed Training}
\label{sec:sgd_baselines}

\chadded{Next, to evaluate the performance of our gradient compressors in the distributed learning environment (as opposed to the gradient reconstruction quality in MSE),} we train a neural network $\net$ (e.g., image classification network) using gradients compressed by the pre-trained \chreplaced{VQ-VAE gradient compressors}{encoder}.
We compare \chreplaced{the learning trajectory -- accuracy of $\net$ as a function of epoch in the distributed learning -- of our VQ-VAE gradient compressors}{it} against the following baseline gradient compression schemes:
\begin{itemize}
\item \textbf{Random-$k$}~\cite{wang2018atomo, koloskova2019decentralizeda}: Instead of communicating the full gradient vector $\mathbf{g} \in \mathbb{R}^d$, we only transmit \chadded{$k$} randomly chosen \chdeleted{$k$} coordinates, thus reducing the communication cost by a factor of $\frac{k}{d}$. 
\item \textbf{Top-$k$}~\cite{shi2019understanding,stich2018sparsified}: Similar to Random-$k$, this approach also communicates $k$ of the $d$ coordinates of a gradient vector and improves the communication cost by a factor of $\frac{k}{d}$. However, this time, we take the top $k$ coordinates with the largest magnitude and discard the remaining $(d-k)$ coordinates.
\item \textbf{QSGD}~\cite{alistarh2017qsgd, basu2019qsparse}: Gradients normally stored using 32-bit floating point numbers are quantized using fewer bits before being transmitted.
\item \textbf{Coordinated Sampling}~\cite{shi2019distributed}:
The nodes \textit{cyclically} select a batch of $k$ coordinates at every iteration and communicate only those $k$ coordinates.
\end{itemize}

All runs were repeated 20 times with different random seeds, and we report average performance along with standard error.

\cref{fig:grad_performance} shows that $\net$ trained with gradients compressed using the distributed VQ-VAE leads to substantial accuracy improvement over the separate VQ-VAE without the extra communication cost. Moreover, we see that VQ-VAE based compression leads to about $3\times$ faster convergence than the baselines (see \cref{fig:grad_comparison_test_acc}).  
\chreplaced{This shows that the correlations between the gradients of two models are \textit{learnable} and a distributed encoder that captures those correlations can be used effectively in distributed learning. We emphasize that we did not modify the training objective of the VQ-VAE for this task; our results show that our VQ-VAE trained to achieve good reconstruction performance of the gradients is an effective means for distributed learning for a different downstream task, such as accuracy}{This shows that a learned distributed encoder can indeed benefit from the complex correlation between gradients distributed optimization}.  

%% file: figures/fig_kitty_plot_and_samples.tex
\begin{figure*}[htbp]
\centering
\subfloat[\label{fig:kitti_bpp_psnr}]{\includegraphics[width=0.47\linewidth]{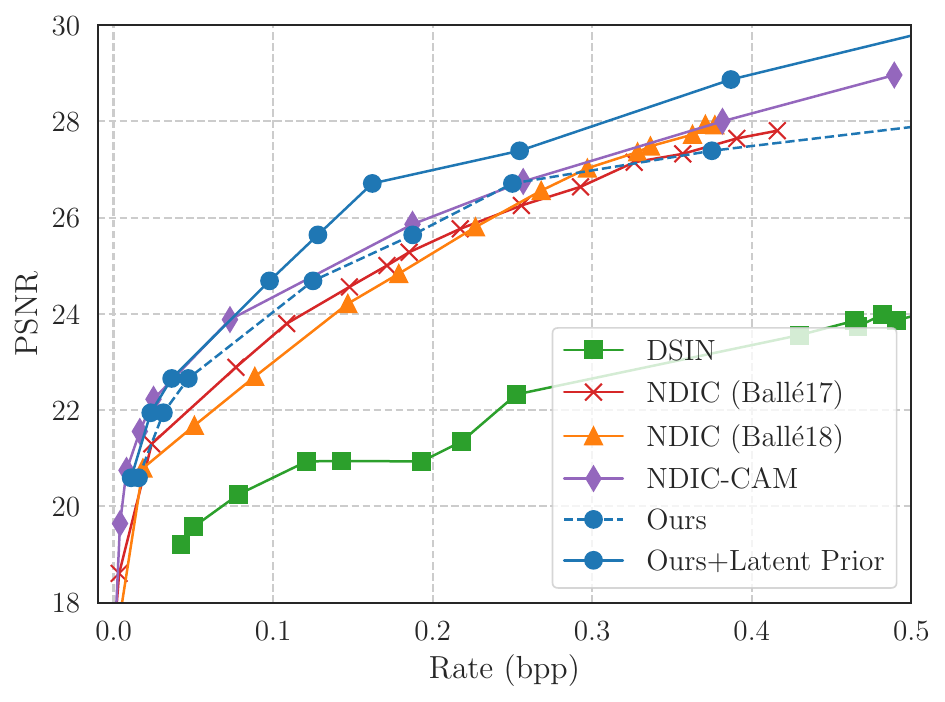}}
\subfloat[\label{fig:kitti_bpp_ssim}]{\includegraphics[width=0.485\linewidth]{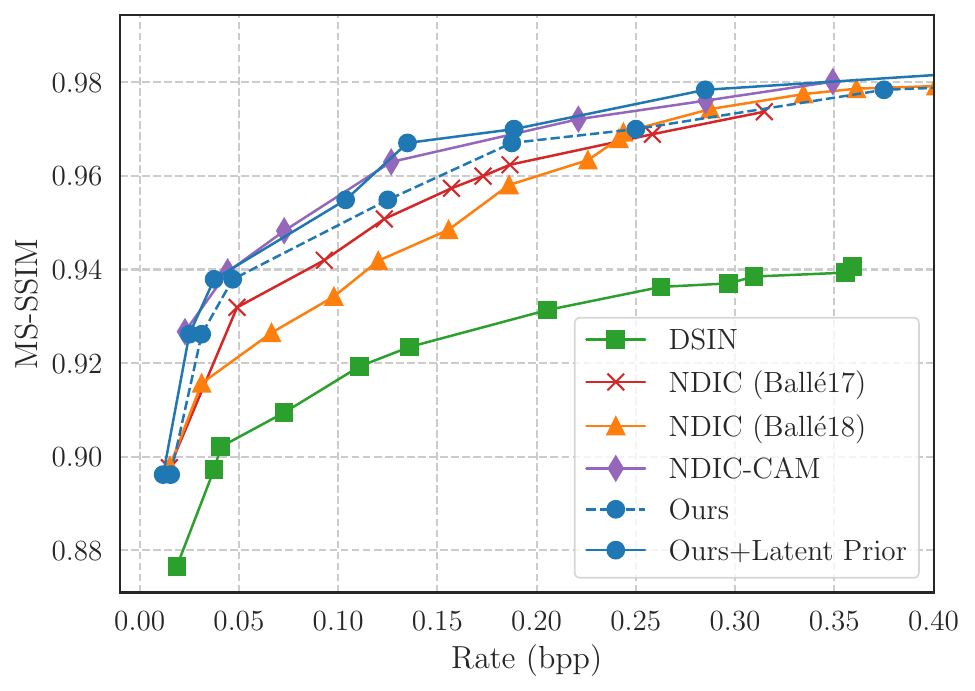}}
\caption{
We compare our method with other DSC methods on the KITTI Stereo image dataset. \textbf{(a)} Comparison of the PSNR rate-distortion curves. \chreplaced{With the latent prior model, we match the previous state-of-the-art method (NDIC-CAM) for low bpp and outperform all methods at higher bpp. }{Our method achieves higher PSNR than other methods for low rates and remains competitive for higher rates. With the latent prior model, we outperform all methods} \textbf{(b)} Comparison of the MS-SSIM rate-distortion curves. \chreplaced{Our method with the latent prior model is able to match the performance of NDIC-CAM.}{Our method achieves higher PSNR than other methods for low rates and remains competitive for higher rates. With the latent prior model, we outperform all methods.}
}
\label{fig:kitti_bpp}
\end{figure*}

%% file: figures/table_param_efficiency.tex
\begin{table}[ht]
\centering
\caption{Number of model parameters for the rate-distortion points (both VQ-VAE and latent prior model) in \cref{fig:kitti_bpp_psnr}.}
\vspace{1em}
\begin{tabular}{@{}lcc@{}}
\toprule
                & Parameter Count (M)   & Factor            \\
\midrule
Ours            & $3.9 - 7.4$           & $1.0$-$1.9\times$ \\
NDIC (Ballé17)  & $16.3$                & $4.2\times$       \\
NDIC (Ballé18)  & $25.0$                & $6.4\times$      \\
NDIC-CAM        & $29.7$                & $7.6\times$      \\
\bottomrule
\end{tabular}
\label{table:param_efficiency}
\end{table}

%% file: figures/fig_celeba_rate_distortion.tex
\begin{figure}[ht]
\centering
\includegraphics[width=0.7\linewidth,]{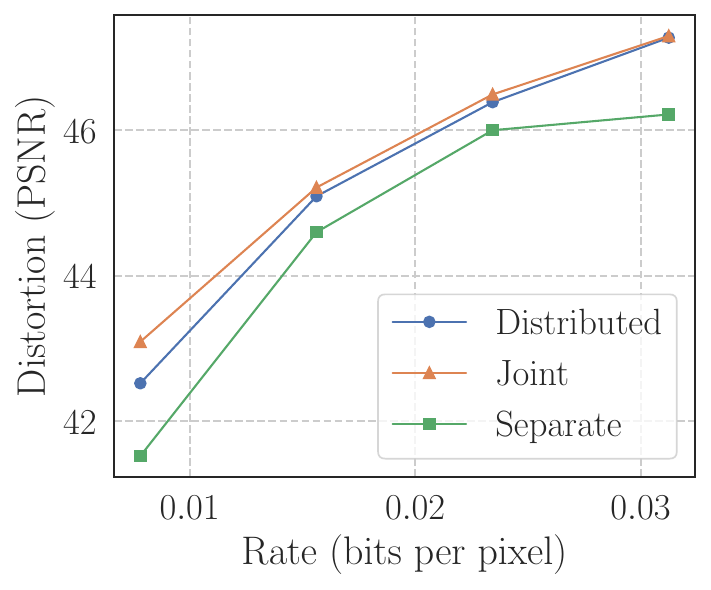}
\caption{Comparison of different coding schemes on CelebA-HQ. This shows that the distributed encoder is able to nearly match the performance of joint encoder, while achieving a noticeable improvement in rate-distortion over the ``separate'' encoder that does not utilize side information.}
\label{fig:celeba_rate_distortion}
\end{figure}


%% file: figures/fig_effect_sideinfo.tex
\begin{figure}[htbp]
\centering
\includegraphics[width=\linewidth]{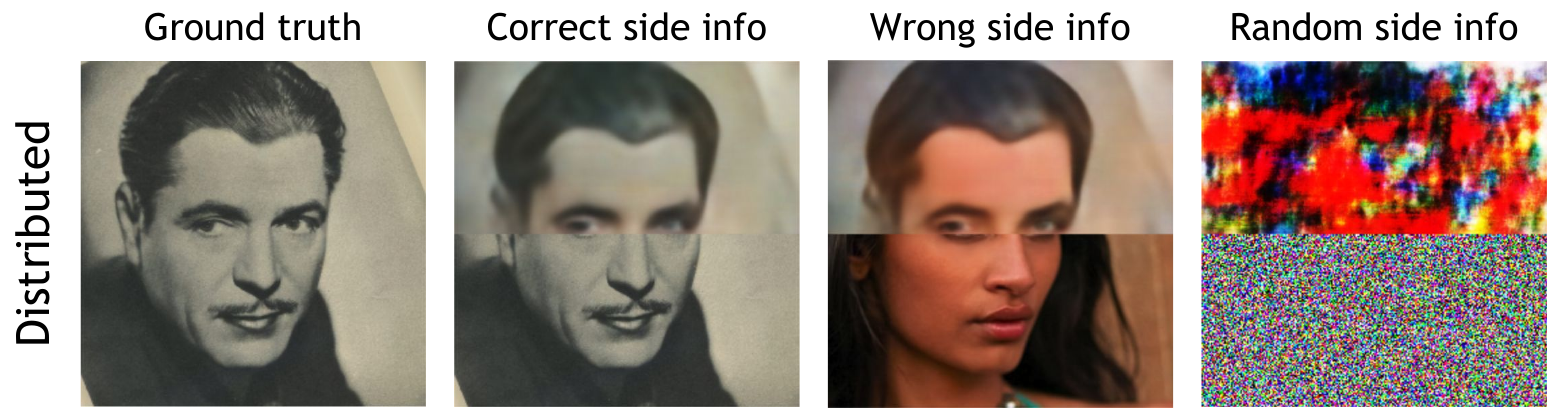}
\caption{Reconstructions from \chreplaced{our distributed VQ-VAE with different types of }{distributed encoder under different} side information.  We can see that providing wrong or random side information to the distributed \chreplaced{VQ-VAE }{decoder} affects the output in a semantic way (e.g. the skin tone changes, while the background remains identical).}
\label{fig:effect_sideinfo}
\end{figure}

%% file: figures/fig_all_grad_plots.tex
\begin{figure*}[htbp]
\centering
\subfloat[\label{fig:grad_vqvae_mse}]{\includegraphics[width=0.33\linewidth]{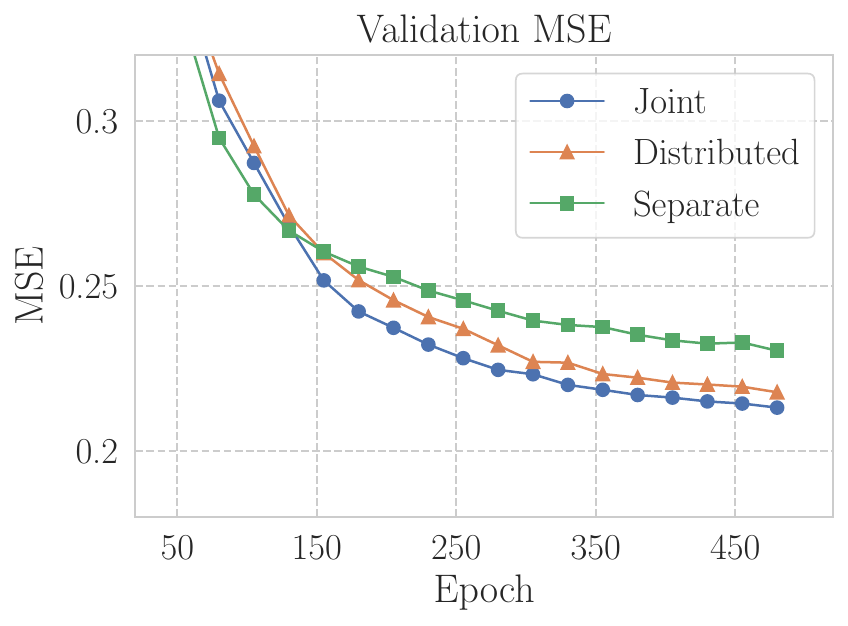}}
\subfloat[\label{fig:grad_performance}]{\includegraphics[width=0.33\linewidth]{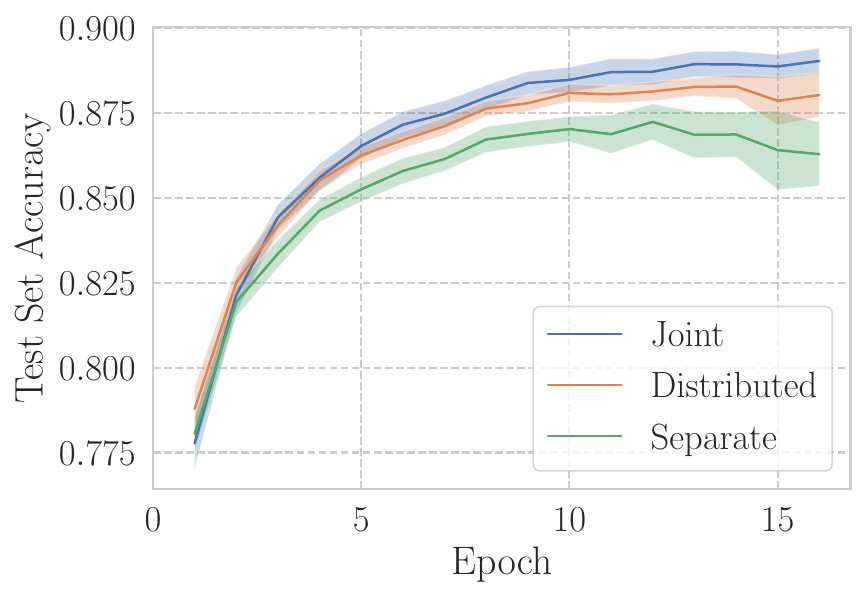}}
\subfloat[\label{fig:grad_comparison_test_acc}]{\includegraphics[width=0.33\linewidth]{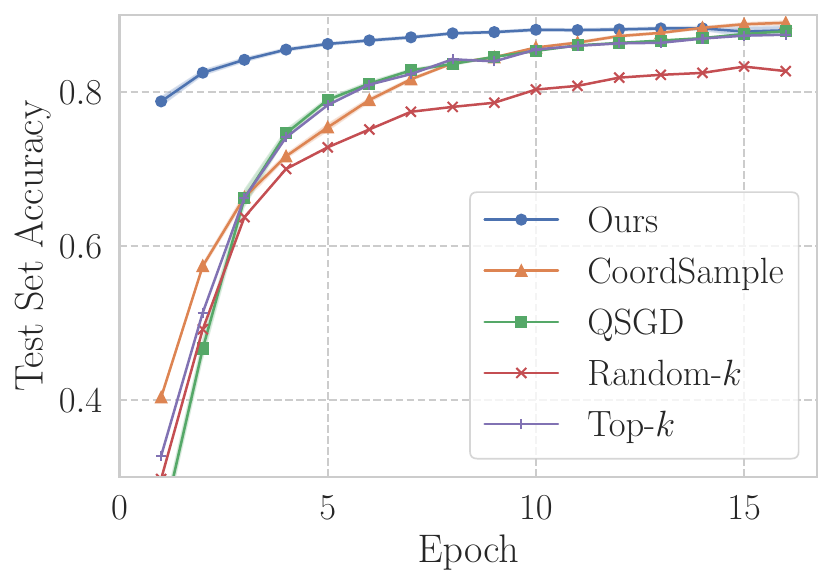}}
\caption{
Our results show that our distributed VQ-VAE is an effective method for gradient compression in the distributed learning setting. \textbf{(a)} Average $\ell_2$ distortion over the validation split during the course of VQ-VAE training.  We observe that the distributed encoder significantly outperforms the separate encoder and approaches the distortion of the joint encoder. \textbf{(b)} Plot of the average test set accuracy over the course of training $\net$ using compressed gradients. We observe that the distributed encoder leads to comparable accuracy as the joint encoder, but using about only half of the communication cost. Shaded regions represent standard error.
\textbf{(c)} Distributed training performance with compressed gradients using a distributed encoder in comparison to other baseline methods in terms of test set accuracy.  Note that standard error is very small and barely visible, and may require zooming in on an electronic copy of this manuscript.
}
\label{fig:all_grad_plots}

\end{figure*}

%% file: 4discussion.tex
\section{Discussion and Analysis}
\label{sec:discussion}

In this section, we further provide a discussion and analysis of our framework. Namely, we first analyze the effect of reconstruction diversity as we vary side information. Next, we perform a sanity check to see whether our approach is indeed performing DSC by comparing it with the optimal distributed compressor. Finally, we compare the performance of our approach with DISCUS \cite{DISCUS1999} in a setting with a simpler correlation structure between the source and side information.

\subsection{Role of Side Information}
In \cref{sec:motivating_example}, we saw that grouping symbols with maximal distance from each other in the same group allows the decoder to determine the correct symbol using the side information. Here, we investigate to what extent our models perform such a grouping (also known as \textit{binning}). If approximate binning were occurring, the same codeword (i.e., bin index) would be decoded into different images for different side information. In other words, a distributed compressor should have high reconstruction diversity for a single codeword as we vary the side information. We refer to this as \textit{bin diversity}.

On the other hand, a model that does not perform binning should decode a single codeword to similar symbols regardless of what side information is given to the decoder.  To test this hypothesis, we train two different models:
\textbf{Ours} was trained with the correct side information as was done for other experiments, and 
\textbf{Uncorrelated SI} was trained by replacing the side information with random ones from other irrelevant samples in the dataset, thus making the input and side information completely independent.  We expect this model to ignore side information and achieve lower bin diversity.

To quantitatively measure diversity, we used average pairwise distance with respect to $\ell_2$ norm and LPIPS distance \cite{zhang2018unreasonable}, which has been used in the literature \cite{srivastava2017veegan,metz2016unrolled} as a measure of sample diversity.  Both metrics were computed and averaged over all images in the CelebA-HQ test dataset.

As shown in \cref{table:diversity}, the VQ-VAE trained with correlated side information exhibits much higher bin diversity in both metrics. In other words, the images within each bin are much farther from each other compared to the other model, suggesting that some form of approximate binning is happening.  The uncorrelated VQ-VAE has particularly low bin diversity with respect to the LPIPS distance, meaning there is very little perceptual difference regardless of what side information is used.

\input{figures/table_diversity}

\subsection{Performance of the Learned Compressor on Synthetic Sources.}
Another natural question is whether our \textit{learned} distributed compressor is performing DSC and strictly better than the joint compressor. 
We investigate this using a synthetic data source studied in \cite{DISCUS1999} with two correlated Gaussian sources: $Y = X + N$, $X \sim \mathcal{N}(0, 1)$, and $N \sim \mathcal{N}(0, 0.1^2)$.
The asymptotically optimal (i.e.,~in the limit of compressing infinitely many symbols together) rate-distortion curve of these sources is known analytically, so we can check how our model compares to the theoretical limit.

\cref{fig:gaussian_rate_dist} shows that both learned methods significantly outperform the asymptotically optimal encoder without side information (SI) for low rates. This is \chdeleted{a} concrete evidence that our learned encoder is actually performing DSC (as opposed to simply achieving a very good compression rate for single source coding). Moreover, the distributed compression performance remains very close to that of joint compression \chadded{depicted in \cref{fig:dist-vs-joint}}, showing the efficacy of our practical distributed coding scheme.

At higher rates, however, the value of side information quickly diminishes, and the learned methods perform worse than even the encoder without SI. This is expected as the optimal curves presented in this figure are asymptotic in the limit of jointly compressing infinitely many symbols, which is clearly not the case for the learned methods that compress each symbol one at a time. 

We also note that a recent work by Ozyilkan, Ball\'e, and Erkip has shown that the distortion of learning-driven DSC can be further improved by explicitly optimizing the entropy and tailoring the design to scalar sources such as Gaussian and Laplacian~\cite{ozyilkan2023neural}. Interestingly, they show that neural distributed compressors bin their source data without explicitly training for it and 
can nearly attain the asymptotic Wyner-Ziv rate-distortion bound for scalar Gaussian and Laplacian sources. 
In contrast, our architectural design and training methodologies are tailored for high-dimensional data sources with complex distributions (but still utilize side information for Gaussian sources as shown in Figure~\ref{fig:gaussian_rate_dist}).

\input{figures/fig_optimality_and_quantization}

\subsection{Comparison with Non-learning Baselines} \label{sec:nonlearning}
\subsubsection{Setup} 
We consider a simple but systematic correlation structure: correlated i.i.d. Bernoulli sources~\cite{DISCUS1999}.  
Specifically, we consider a pair of source sequence and side information $(\mathbf{x,y}) = \{x_i,y_i\}_{i=1}^{648}$, where $y_i \oplus x_i$ is an i.i.d. Bernoulli random variable.

We compare the performance of our Neural DSC approach\footnote{Implementation details are provided in Appendix ~\ref{sec:ndsc-iid}.} with the Distributed Source Coding Using Syndromes (DISCUS), a non-learning based constructive scheme that is designed specifically for this setup and shown to be near optimal~\cite{DISCUS1999}.  
Our experimental results are shown in \cref{fig:iid_nonlearning}. As expected, our neural DSC performs worse than DISCUS (labeled as LDPC); however, the performance of distributed neural DSC is comparable with the performance of joint source coding, which serves as a sanity check. 

\subsubsection{Interpretation} Unlike other experiments, here we focus on a systematic correlation structure, where the problem at hand is learning a complex {\em algorithm}; there is no complexity in the data distribution or correlation structure. In \cref{fig:iid_nonlearning}, we plot the block error rate of our method in the joint and distributed cases and compare it with a DISCUS baseline. We implemented the DISCUS~\cite{DISCUS1999} using a rate $3/4$ Low-Density Parity Check (LDPC) code, specified in IEEE 802.11~\cite{standard:ldpc},~\cite{matlab:ldpc}, which maps 486 message bits to a length-648 codeword. The parity-check matrix is created using the ldpcQuasiCyclicMatrix function, with the block size 27 and prototype matrix P provided in Appendix~\ref{sec:discus_details}.

These results show that even in such scenarios, our method learns a nontrivial compression algorithm (i.e., a compression scheme without side information achieves the block error rate $\approx 1$). We note, however, that our neural DSC results in higher error rates than the DISCUS algorithm. This observation suggests that learning DISCUS, or an alternative method that outperforms DISCUS, is a highly non-trivial task. This is perhaps not surprising given that learning channel codes using neural networks is, in general, highly challenging~\cite{TurboAE,KOCodes,ProductAE} and typically requires a tailored selection of neural architectures and training methodologies, which is beyond the scope of this paper.

\input{figures/fig_iid.tex}

%% file: figures/table_diversity.tex
\begin{table}[htbp]
\centering
\caption{Diversity of decoded samples as we vary side information.}
\vspace{1em}
\begin{tabular}{@{}lcc@{}}
\toprule
                 & Diversity (LPIPS) & Diversity ($\ell_2$) \\ \midrule
Ours             & 0.1130            & 35.80 \\
Uncorrelated SI  & 0.00469           & 4.167 \\
\bottomrule
\end{tabular}
\label{table:diversity}
\end{table}

%% file: figures/fig_optimality_and_quantization.tex
\begin{figure}[htbp]
\centering
\includegraphics[width=0.7\linewidth]{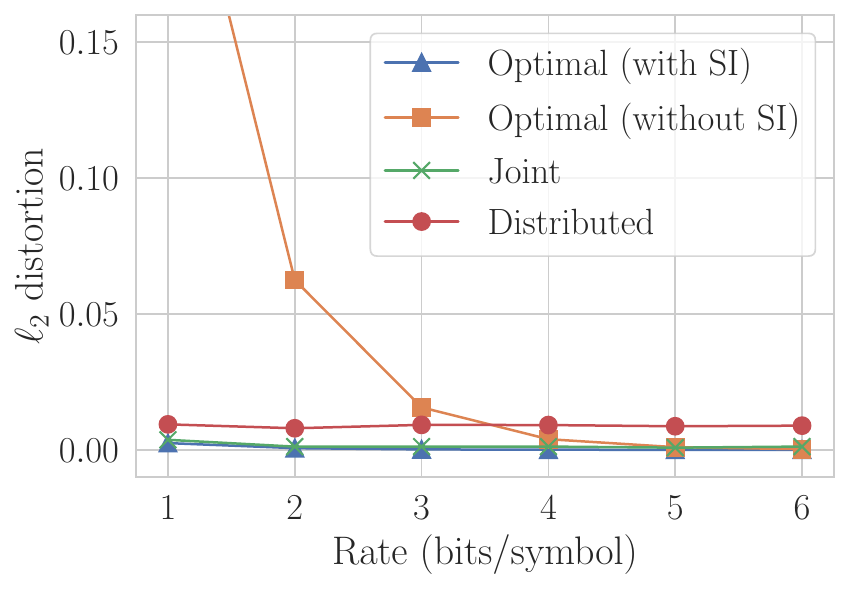}
\caption{Rate-distortion points of learned joint and distributed encoders compared to theoretically best compressors. The ``Optimal'' curves represent the asymptotically optimal rate-distortion points and are included as reference.  Note that these are theoretical limits and do not correspond to any concrete compression scheme.}
\label{fig:gaussian_rate_dist}
\end{figure}

%% file: figures/fig_iid.tex
\begin{figure}[htbp]
\centering
\includegraphics[width=0.7\linewidth]{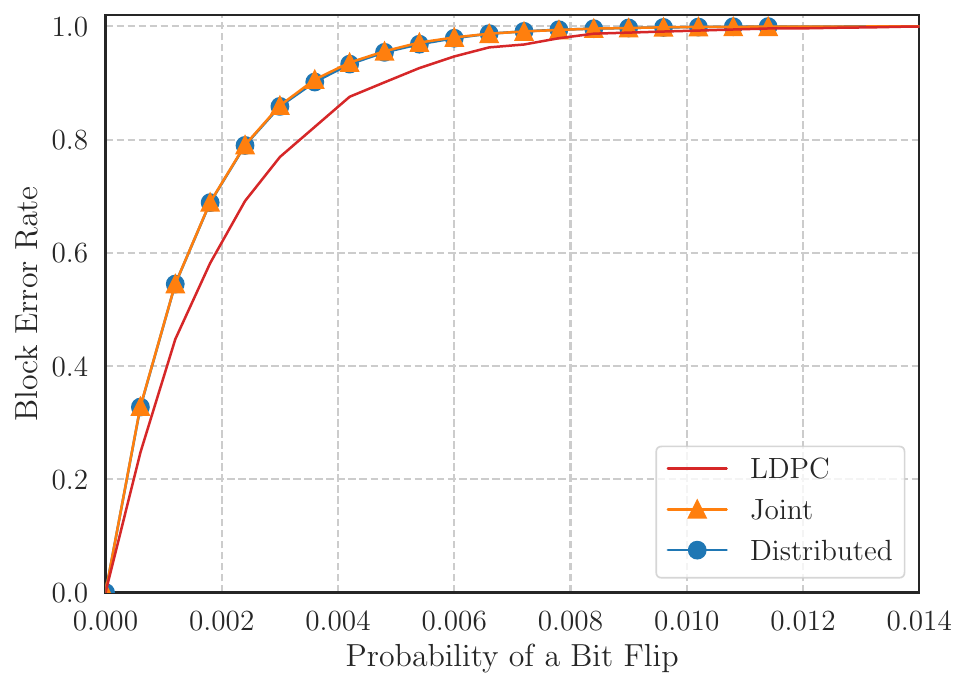}
\caption{Block error rate for our VQ-VAE architecture and an LDPC implementation of DISCUS on i.i.d. Bernoulli messages.} 
\label{fig:iid_nonlearning}
\end{figure}

%% file: 5conclusion.tex
\section{Conclusion}
We presented Neural DSC, a framework for distributed compression of correlated sources. Our method is built on the power of modern deep generative models, namely VQ-VAEs, which are excellent data-driven models to represent high-dimensional distributions using compressed discrete latent codes. Our training objective is justified in part by the connection we establish between distributed source coding and the modified ELBO for asymmetric VAEs.

Empirically, we show that our method achieves the new state-of-the-art performance in terms of PSNR on stereo image compression. Our approach remains competitive with the current best results for MS-SSIM. We also show that our method is able to leverage complex correlations far beyond spatial similarity for better compression performance, approaching the joint compression rate. Finally, we show that our method is not limited to images and show a promising proof of concept for compressing gradients for distributed training. 

We believe our work provides encouraging initial results for practical data-driven DSC. 
We hope to see further developments that will bring insights on DSC pioneered by Berger, Tung, Slepian, and Wolf, and further enriched by decades of research in information theory, to practical applications -- with the aid of learning and a data-driven approach. 
 
Some specific future directions include exploring a hybrid approach that combines the DISCUS algorithm with neural non-linear representation learning \chdeleted{would be interesting}. Such an integration could harness both methodologies' strengths while simplifying the overall complexity. Additionally, our research could be extended to scenarios involving more than two distributed encoders collaborating to compress correlated data for a single decoder. Furthermore, we encourage exploring scenarios where the decoder's objective shifts from reconstructing the raw correlated sources to reconstructing a function derived from these correlated sources. This scope expansion will open new avenues for applying DSC to a broader range of real-world problems.

%% file: 6appendix.tex
\appendices
\section{Architectural Details}

\subsection{VQ-VAE Architecture}
\label{app:vq_vae_arch}

For our image compression experiments, we used a convolutional VQ-VAE architecture similar to the one used in \cite{ramesh2021zero} with residual connections.

Both the input and the side information to this network have the shape $\chadded{3 \times }128 \times 256$ (i.e., vertical halves of a full $\chadded{3 \times }256 \times 256$ image).
The encoder scales down the input image by a factor of $4\times$ or $8\times$ both vertically and horizontally, producing latent variables of shape $32 \times 64$ and $16 \times 32$, respectively. Each dimension of the latent variable is allotted different numbers of bits (i.e.~codebook bits), which range from $1$ to $8$ in our experiments. The decoder conversely takes a discrete latent variable and upscales it by a factor of $4\times$ or $8\times$ to produce a reconstruction of the original shape.

A detailed specification of the architecture is provided in \cref{fig:vqvae_arch_conv}.  The notation (\texttt{Tconv}) ``\texttt{Conv} $A \times B$ ($C \to D$)'' represents a 2D (transposed) convolution with kernel size $A$ and stride $B$ with input and output channels $C$ and $D$, respectively.  The boxes ``\texttt{Residual Block} ($A\to B$)'' represent a two-layer residual network.  We used GELU (Gaussian Error Linear Unit) activation for all layers except for the very last convolution of the decoder, for which we used sigmoid. \chadded{$\oplus$ represents the channel-wise concatenation of the image representations.} The exact number of channels may differ for different rate-distortion points, and we refer the reader to the supplementary code submission for full details.

\input{figures/fig_vqvae_arch_conv}

We used a small network (denoted ``\texttt{SI Net}'') shared by the encoder and decoder to preprocess the side information.  Whenever the encoder or decoder does not receive side information (for distributed and separate VQ-VAEs), we simply replace the output of \texttt{SI Net} with a zero tensor of the same shape. \chadded{On the decoder side, the output of the \texttt{SI Net} is passed through the \texttt{Cond Net}; the architecture of this network is nearly identical to the encoder. The architecture of the \texttt{SI Net} and the \texttt{Cond Net} are detailed in \cref{fig:si_net_arch_conv}.}

\input{figures/fig_si_net_arch_conv}

We note that all three VQ-VAE variants have the same architecture and number of parameters for the auto-encoder portion (the portion that the horizontal line goes through in \cref{fig:vqvae_arch_conv}). Thus, there is no architectural advantage among the VQ-VAE variants.

\subsection{Specifying the Target Rate}

The shape and the number of codebook bits determine the total rate.
This is different from neural compression models that use a regular VAE, which controls the rate-distortion by reweighting the training objective (ELBO). As mentioned in \cref{sec:our_method}, the fact that we have a hard constraint on the rate is useful when working with a hard communication limit.

There are several ways to achieve a desired target rate.  For example, an $8\times$ downscaling encoder with 4 codebook bits produces a compressed message of size $16 \times 32 \times 4 = 2048$ bits. The same rate can be achieved using a $4\times$ downscaling encoder with 1 codebook bit: $32 \times 64 \times 1 = 2048$ bits.  This flexibility is what allows us to have a fairly granular control over the target rate at the cost of hyperparameter choices.

\textbf{Hyperparameters for KITTI Stereo Experiment. }
Due to the above choices, we use differently sized networks for each rate-distortion curve for our stereo image compression experiments. While the model definition is included in the code, we include the relevant hyperparameter information in \cref{table:kitti_hparams} below.

\input{figures/table_kitti_hparams}

\textbf{Hyperparameters for CelebA-HQ Experiment. }
For this experiment, we used the single $8\times$ downscaling encoder with codebook bits $\{1, 2, 3, 4\}$, resulting in the total rates of $\{512, 1024, 1536, 2048\}$.

\textbf{Distributed optimization. }
For gradient compression VQ-VAEs, we followed the same architecture as the image compression experiments but replaced all convolutional layers with fully-connected layers. Full specification of the network is available at \href{https://github.com/acnagle/neural-dsc}{\texttt{https://github.com/acnagle/neural-dsc}}.

\section{Stereo Image Dataset}

For our stereo image compression experiment, we follow the setup of \cite{NDIC}.  First, we construct training/test datasets using the files specified in the official repository for \cite{ayzikA2020dsin} (\url{https://github.com/ayziksha/DSIN/tree/master/src/data_paths}).
Then we apply the same preprocessing steps of \cite{NDIC}, where we first take the center crop of size $370 \times 740$, then resize it to $128 \times 256$ using PyTorch transformations. This results in the training split containing 1576 pairs of stereo images and the test split containing 790 pairs. \chadded{In \cref{fig:kitti_bpp_ldmic}, we show the performance of LDMIC \cite{zhang2023ldmic} with two sources against all other methods. 
For LDMIC, we vary the compression rate for one of the two views, denoted by $\bm{x}$, (Rate (bpp) on the x-axis) and fix the compression rate for the other one, denoted by $\bm{y}$, as 24 bits-per-pixel so that LDMIC can losslessly compress the second view $\bm{y}$. 
In principle, the LDMIC for two image views should recover our setup.  However, the results in \cref{fig:kitti_bpp_ldmic} demonstrate that the performance of LDMIC when configured in this way is generally much worse than our approach and other neural distributed source coding schemes designed for compression with side information.
}

\input{figures/fig_sample_stereo_vs_face}

\input{figures/fig_kitti_with_ldmic}

\section{Training Details}
\label{sec:dis_training}
We trained our models for up to 1000 epochs on a DGX machine for stereo image compression experiments.  Some training runs were early stopped because the validation performance started to plateau.
For CelebA-HQ experiments, we trained the VQ-VAEs for a total of 20 epochs distributed over two Nvidia GTX 2080 GPUs.
We trained the fully connected VQ-VAEs for 500 epochs on a single GPU for gradient compression experiments.

We evaluated validation loss after each epoch in all cases and observed no overfitting.  This leads us to believe that it may be possible to further improve the performance of our method by training a larger network, which we leave for future work.

The latent prior models were implemented as an autoregressive transformer decoder network. All latent prior models were trained with an initial learning rate of 3e-4 with a cosine annealing learning rate decay. Training lasted 100 epochs and the learning rate had a linear warmup period for the first 2000 steps of training. The Adam optimizer was used with parameters $\beta_1 = 0.9$, $\beta_2 = 0.999$, $\epsilon = 1e-8$. The hyperparameters for each latent prior model are provided in \cref{table:prior_hparams}. The models were trained on an RTX 3090 GPU, and due to memory constraints some of the latent prior models had to be trained with fewer blocks, fewer attention heads, and a smaller batch size.

\input{figures/table_prior_hparams}

\subsection{NDIC Training Details}
For the CelebA-HQ compression experiment, we trained NDIC \cite{NDIC} using the official code released by the authors. We used the ``Balle18'' \cite{balle2018variational} backbone and trained the model for 10 epochs ($\approx$ 300K examples). We used the model checkpoint with the best validation set performance for evaluation. While \cite{NDIC} report training with a batch size of 1 for 500K steps, we chose to use a batch size of 20. This was done for several reasons. First, using a batch size of 1 is very inefficient as we do not benefit from GPU parallelism. It also leads to high variance in the gradient, often leading to slower convergence. In our training, the loss plateaued well before reaching 300K total examples. 

\subsection{NDIC-CAM Training Details}
\label{app:ndic_cam}
We report the hyperparameters used to train the NDIC-CAM method on the KITTI Stereo dataset while optimizing for PSNR. All models were trained for 1000 epochs, at which point the models were well-converged. The hyperparameters and their corresponding results are shown in \cref{table:ndic_cam_hparams}. The results shown for MS-SSIM are reported from the NDIC-CAM paper \cite{NDIC-CAM}.

\input{figures/table_ndic_cam_hparams}

\subsection{LDMIC Training Details}
\label{app:ldmic}
\chadded{We train all models following the hyperparameters used in the paper which proposes the LDMIC framework. For MSE distortion, we train a model for each $\lambda \in \{2,4,8,16,32,64,128,256,512,1024,2048\}$ with an initial learning rate $10^{-4}$ and train for 400 epochs while decaying the learning rate by a factor of two for every 100 epochs of training. We use these models' weights to initialize the models' weights trained with MS-SSIM distortion and train for an additional 400 epochs with an initial learning rate $5\times 10^{-5}$. We use a batch size of 32 for all training runs.}

\subsection{Neural DSC Training Details for i.i.d. Bernoulli experiments}
\label{sec:ndsc-iid}
A schematic of the VQ-VAEs trained in the joint and distributed settings is shown in \cref{fig:vqvae_arch_discus}. A primary difference between this architecture and the architecture for the other experiments is that the output of the decoder of this architecture has a learnable parameter $\alpha$, which is used to interpolate the reconstructed sequence and the side information linearly, i.e., the output of the network is $(1 - \alpha)\reconst + \alpha \sinfo$, where $\reconst$ is the reconstruction and $\sinfo$ is the side information. This enables the VQ-VAE architecture to find a tradeoff between the reconstruction and the side information as the final reconstruction of the Bernoulli sequence. Our VQ-VAE architectures were trained on i.i.d. Bernoulli sequences of length 648, and the block error rate was computed on a fixed validation set of 10,000 i.i.d. Bernoulli sequences. All models were trained with the same hyperparameters: 128 batch size, 3e-4 learning rate, 0.15 commitment cost (for the VQ-VAE loss), and 10 epochs. The Adam optimizer was used with parameters $\beta_1 = 0.9$, $\beta_2 = 0.999$, $\epsilon = 10^{-8}$. 

\input{figures/fig_vqvae_arch_discus}

\section{DISCUS Details}\label{sec:discus_details}

For the experiment on the correlated i.i.d. Bernoulli sources (\cref{fig:iid_nonlearning}), we used the following prototype matrix P for the quasi-cyclic LDPC code. 


\[
P^\top = \begin{bmatrix}
 16 &  25 &  25 &   9 &  24 &   2 \\
 17 &  12 &  18 &   7 &   5 &   2 \\
 22 &  12 &  26 &   0 &  26 &  19 \\
 24 &   3 &  16 &   1 &   7 &  14 \\
  9 &   3 &  22 &  17 &   1 &  24 \\
  3 &  26 &  23 &  -1 &  -1 &   1 \\
 14 &   6 &   9 &  -1 &  -1 &  15 \\
 -1 &  21 &  -1 &   7 &  15 &  19 \\
  4 &  -1 &   0 &   3 &  24 &  -1 \\
  2 &  15 &  -1 &  -1 &  15 &  21 \\
  7 &  22 &   4 &   3 &  -1 &  -1 \\
 -1 &  -1 &  -1 &  23 &   8 &   2 \\
 26 &  15 &   4 &  -1 &  -1 &  -1 \\
 -1 &  -1 &  -1 &  16 &  13 &  24 \\
  2 &   4 &   8 &  -1 &  -1 &  -1 \\
 -1 &  -1 &  23 &  -1 &  13 &   3 \\
 21 &  -1 &  11 &  21 &  -1 &  -1 \\
 -1 &  16 &  -1 &  -1 &  11 &   2 \\
  1 &  -1 &  -1 &   0 &  -1 &   1 \\
  0 &   0 &  -1 &  -1 &  -1 &  -1 \\
 -1 &   0 &   0 &  -1 &  -1 &  -1 \\
 -1 &  -1 &   0 &   0 &  -1 &  -1 \\
 -1 &  -1 &  -1 &   0 &   0 &  -1 \\
 -1 &  -1 &  -1 &  -1 &   0 &   0 \\
\end{bmatrix}.
\]

%% file: figures/fig_vqvae_arch_conv.tex
\begin{figure*}[!h]
\centering
\includegraphics[width=1.0\textwidth]{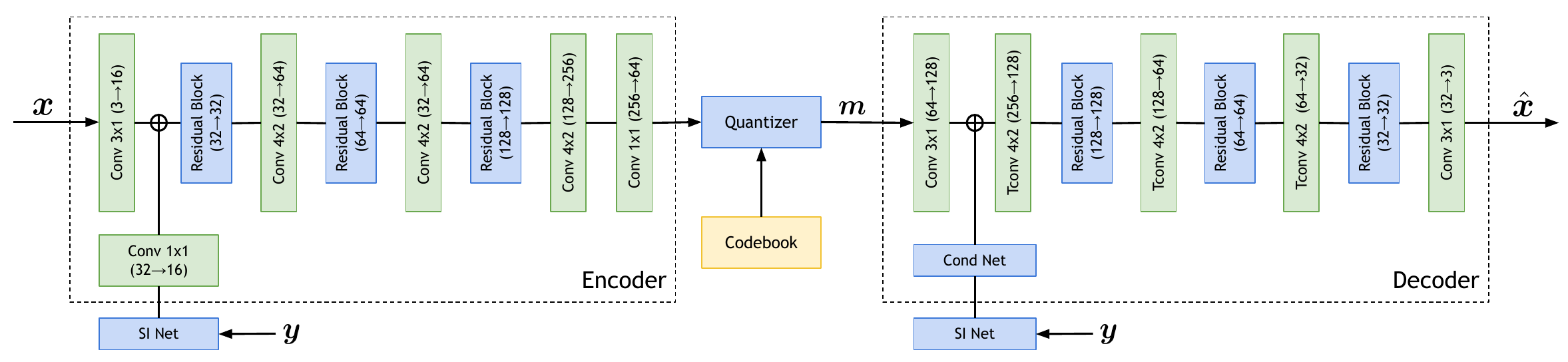}
\caption{Conditional VQ-VAE Architecture used for the image compression experiments.}
\label{fig:vqvae_arch_conv}
\end{figure*}

%% file: figures/fig_si_net_arch_conv.tex
\begin{figure*}[!h]
\centering
\includegraphics[width=1.0\textwidth]{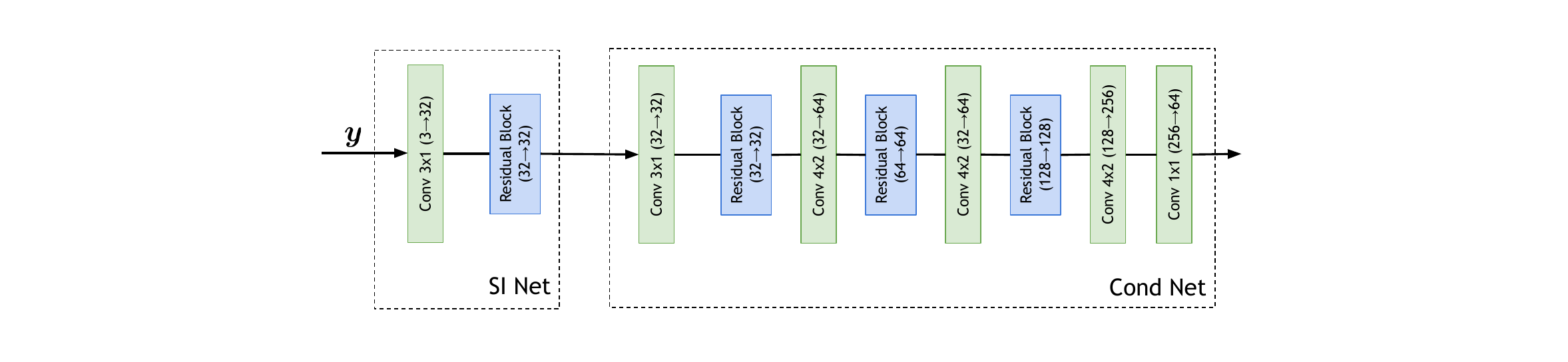}
\caption{Architecture for the \texttt{SI Net} and \texttt{Cond Net} used for the image compression experiments. For the gradient compression and synthetic data compression experiments, the \texttt{SI Net} and \texttt{Cond Net} share a similar architecture as their respective counterparts in the image compression setup, including the same number of layers, but the convolutions are replaced with fully connected layers.}
\label{fig:si_net_arch_conv}
\end{figure*}

%% file: figures/table_kitti_hparams.tex
\begin{table*}[ht]
\centering
\caption{\footnotesize Different models and their associated hyperparameters for the stereo image compression experiment. Each row represents a single rate-distortion point in \cref{fig:kitti_bpp_psnr}.}
\begin{tabular}{@{}cccccc@{}}
\toprule
Downscaling Factor & Codebook Size (bits) & Rate (bpp)   & PSNR & MS-SSIM & Parameter count \\
\midrule
$8\times$   & 1     & 0.0156    & 20.77     & 0.867     &  3,954,853 \\
$8\times$   & 2     & 0.0312    & 21.884    & 0.902     &  3,955,111 \\
$8\times$   & 3     & 0.0469    & 22.658    & 0.917     &  6,250,747 \\
$8\times$   & 8     & 0.125     & 24.687    & 0.945     &  4,037,827 \\
$4\times$   & 3     & 0.1875    & 25.642    & 0.959     &  6,276,619 \\
$4\times$   & 4     & 0.2500    & 26.582    & 0.969     &  6,277,651 \\
$4\times$   & 6     & 0.3750    & 27.531    & 0.973     &  6,283,843 \\
$2\times$   & 3     & 0.75      & 28.870    & 0.976     &  2,463,883 \\
$2\times$   & 4     & 1.0       & 30.101    & 0.983     &  2,464,915 \\
\bottomrule
\end{tabular}
\label{table:kitti_hparams}
\end{table*}

%% file: figures/fig_sample_stereo_vs_face.tex
\begin{figure}[htbp]
\centering
\includegraphics[width=0.8\linewidth]{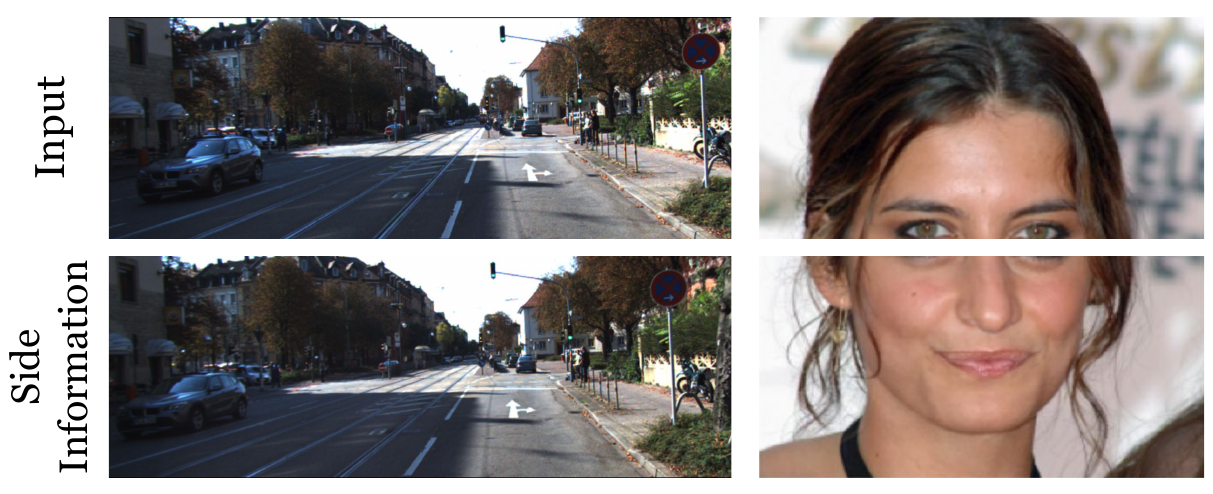}
\caption{\footnotesize \textbf{(Left)} Sample image pair from the KITTI Stereo dataset \cite{geiger2012we} used for the stereo image compression experiment. \textbf{(Right)} Image pairs used to test the model's ability to handle complex correlations beyond spatial similarities. This was obtained by vertically splitting CelebA-HQ \cite{liu2015deep} images.}
\label{fig:sample_stereo_vs_face}
\end{figure}

%% file: figures/fig_kitti_with_ldmic.tex
\begin{figure*}[htbp]
\centering
\subfloat[\label{fig:kitti_bpp_psnr_ldmic}]{\includegraphics[width=0.47\linewidth]{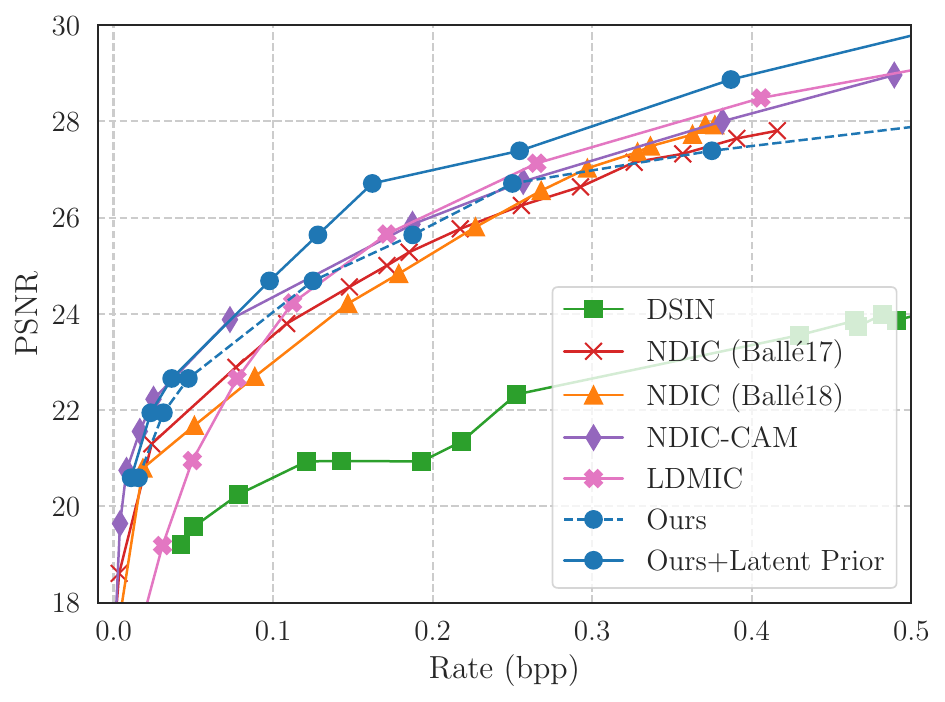}}
\subfloat[\label{fig:kitti_bpp_ssim_ldmic}]{\includegraphics[width=0.485\linewidth]{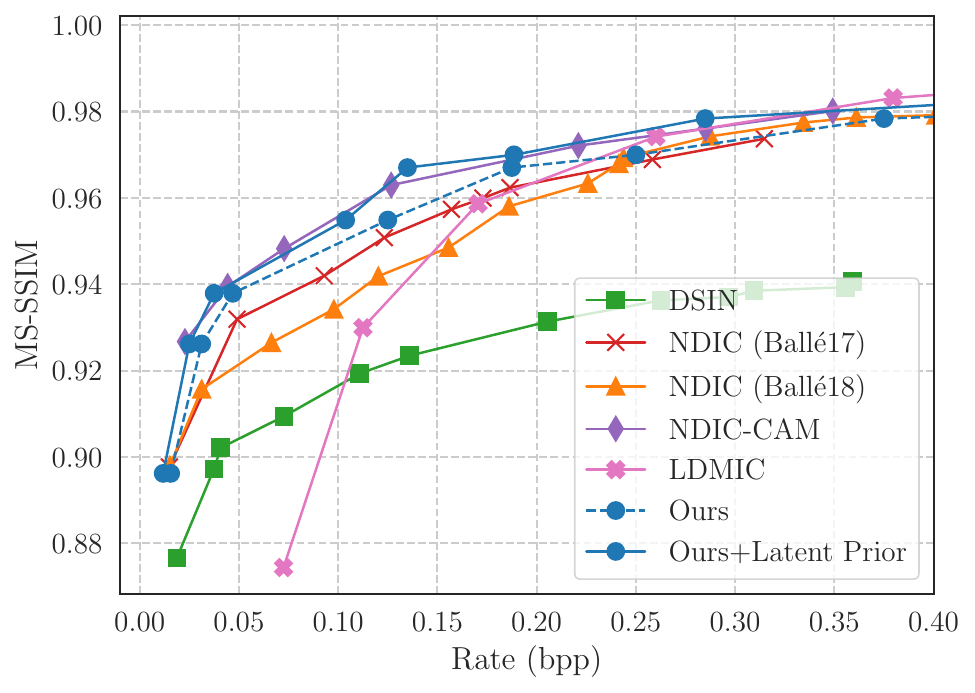}}
\caption{Recreation of Figure 6 from our paper with results from LDMIC \cite{zhang2023ldmic} on KITTI Stereo. We measure the distortion with PSNR \textbf{(a)} and MS-SSIM \textbf{(b)}.}
\label{fig:kitti_bpp_ldmic}
\end{figure*}

%% file: figures/table_prior_hparams.tex
\begin{table*}[ht]
\centering
\caption{\footnotesize Different VQ-VAE models and the associated hyperparameters for their latent prior model counterparts for the stereo image compression experiment. Each row represents a single rate-distortion point in \cref{fig:kitti_bpp_psnr}.}
\begin{tabular}{@{}ccccccc@{}}
\toprule
\thead{Downscaling\\Factor} &
\thead{Codebook Size\\(bits)} &
\thead{Rate (bpp)\\with Latent Prior} &
\thead{\# Attn Heads\\Per Block} &
\thead{\# Blocks} &
\thead{Batch Size} &
\thead{Parameter\\count} \\
\midrule
$8\times$   & 1     & 0.0110    & 4    & 4    & 128    & 859,394   \\
$8\times$   & 2     & 0.0233    & 4    & 4    & 128    & 859,908   \\
$8\times$   & 3     & 0.0364    & 4    & 4    & 128    & 860,936   \\
$8\times$   & 8     & 0.0978    & 4    & 4    & 128    & 924,672   \\
$4\times$   & 3     & 0.1281    & 4    & 4    & 16     & 1,057,544 \\
$4\times$   & 4     & 0.1622    & 4    & 4    & 16     & 1,059,600 \\
$4\times$   & 6     & 0.2545    & 4    & 4    & 16     & 1,071,936 \\
$2\times$   & 3     & 0.3869    & 2    & 2    & 5      & 1,447,432 \\
$2\times$   & 4     & 0.5395    & 2    & 2    & 5      & 1,449,488 \\
\bottomrule
\end{tabular}
\label{table:prior_hparams}
\end{table*}

%% file: figures/table_ndic_cam_hparams.tex
\begin{table}[ht]
\centering
\caption{\footnotesize Hyperparameters used to train NDIC-CAM \cite{NDIC-CAM} on the PSNR objective. Each row represents a single rate-distortion point in \cref{fig:kitti_bpp_psnr}.}
\begin{tabular}{@{}ccccc@{}}
\toprule
\thead{$\alpha$} &
\thead{$\beta$} &
\thead{$\lambda$} &
\thead{Rate (bpp)} &
\thead{PSNR} \\
\midrule
0   & 0     & $3\times 10^{-5}$    & 0.00116    & 17.533 \\
0   & 0     & $6\times 10^{-5}$    & 0.00408    & 19.644 \\
0   & 0     & $1.1\times 10^{-4}$  & 0.00814    & 20.749 \\
0   & 0     & $1.8\times 10^{-4}$  & 0.01649    & 21.558 \\
0   & 0     & $3\times 10^{-4}$    & 0.02505    & 22.227 \\
0   & 0     & $8\times 10^{-4}$    & 0.07297    & 23.882 \\
0   & 0     & $2\times 10^{-3}$    & 0.18738    & 25.862 \\
0   & 0     & $3\times 10^{-3}$    & 0.25674    & 26.752 \\
0   & 0     & $5\times 10^{-3}$    & 0.38164    & 28.002 \\
0   & 0     & $8\times 10^{-3}$    & 0.48933    & 28.963 \\
\bottomrule
\end{tabular}
\label{table:ndic_cam_hparams}
\end{table}

%% file: figures/fig_vqvae_arch_discus.tex
\begin{figure*}[ht]
\centering
\includegraphics[width=1.0\textwidth]{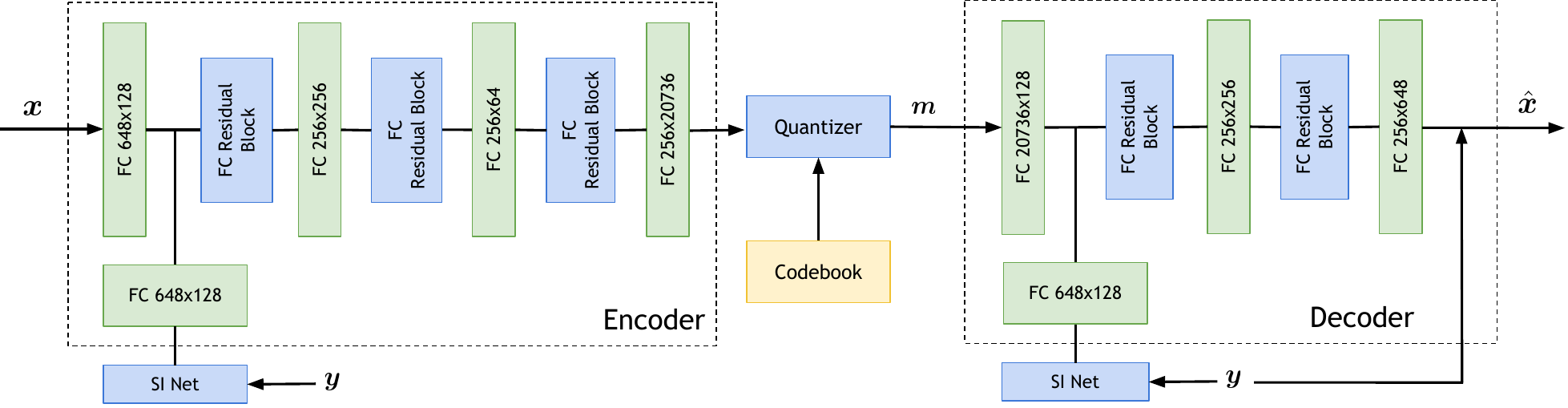}
\caption{Conditional VQ-VAE architecture used for the DISCUS experiments.}
\label{fig:vqvae_arch_discus}
\end{figure*}

%% file: 0main.bbl
\newcommand{\noopsort}[1]{}
\begin{thebibliography}{10}
\providecommand{\url}[1]{#1}
\csname url@samestyle\endcsname
\providecommand{\newblock}{\relax}
\providecommand{\bibinfo}[2]{#2}
\providecommand{\BIBentrySTDinterwordspacing}{\spaceskip=0pt\relax}
\providecommand{\BIBentryALTinterwordstretchfactor}{4}
\providecommand{\BIBentryALTinterwordspacing}{\spaceskip=\fontdimen2\font plus
\BIBentryALTinterwordstretchfactor\fontdimen3\font minus \fontdimen4\font\relax}
\providecommand{\BIBforeignlanguage}[2]{{%
\expandafter\ifx\csname l@#1\endcsname\relax
\typeout{** WARNING: IEEEtran.bst: No hyphenation pattern has been}%
\typeout{** loaded for the language `#1'. Using the pattern for}%
\typeout{** the default language instead.}%
\else
\language=\csname l@#1\endcsname
\fi
#2}}
\providecommand{\BIBdecl}{\relax}
\BIBdecl

\bibitem{Slepian-Wolf1973}
D.~Slepian and J.~Wolf, ``Noiseless coding of correlated information sources,'' \emph{IEEE Transactions on Information Theory}, vol.~19, no.~4, pp. 471--480, 1973.

\bibitem{Cover1975a}
T.~M. Cover, ``\noopsort{a}{A} proof of the data compression theorem of {S}lepian and {W}olf for ergodic sources,'' \emph{{IEEE} Trans. Inf. Theory}, vol.~21, no.~2, pp. 226--228, Mar. 1975.

\bibitem{Berger1978}
T.~Berger, ``Multiterminal source coding,'' in \emph{The Information Theory Approach to Communications}, G.~Longo, Ed.\hskip 1em plus 0.5em minus 0.4em\relax New York: Springer-Verlag, 1978, pp. 171--231.

\bibitem{Tung1978}
S.-Y. Tung, ``Multiterminal source coding,'' {Ph.D.} Thesis, Cornell University, Ithaca, NY, 1978.

\bibitem{Wagner--Kelly--Altug2011}
A.~B. Wagner, B.~G. Kelly, and Y.~Altu\u{g}, ``Distributed rate--distortion with common components,'' \emph{{IEEE} Trans. Inf. Theory}, vol.~57, no.~7, pp. 4035--4057, Jul. 2011.

\bibitem{Wagner--Tavildar--Viswanath2008}
A.~B. Wagner, S.~Tavildar, and P.~Viswanath, ``Rate region of the quadratic {G}aussian two-encoder source-coding problem,'' \emph{{IEEE} Trans. Inf. Theory}, vol.~54, no.~5, pp. 1938--1961, May 2008.

\bibitem{courtade_tsachy_logloss}
T.~A. Courtade and T.~Weissman, ``Multiterminal source coding under logarithmic loss,'' \emph{IEEE Transactions on Information Theory}, vol.~60, no.~1, pp. 740--761, 2014.

\bibitem{Oohama1998}
Y.~Oohama, ``The rate--distortion function for the quadratic {G}aussian {CEO} problem,'' \emph{{IEEE} Trans. Inf. Theory}, vol.~44, no.~3, pp. 1057--1070, 1998.

\bibitem{Berger--Zhang--Viswanathan1996}
T.~Berger, Z.~Zhang, and H.~Viswanathan, ``The {CEO} problem,'' \emph{{IEEE} Trans. Inf. Theory}, vol.~42, no.~3, pp. 887--902, 1996.

\bibitem{Prabhakaran--Tse--Ramchandran2004}
V.~Prabhakaran, D.~N.~C. Tse, and K.~Ramchandran, ``Rate region of the quadratic {G}aussian {CEO} problem,'' in \emph{Proc. {IEEE} Int. Symp. Inf. Theory}, Chicago, IL, June/July 2004, p. 117.

\bibitem{Oohama2005}
Y.~Oohama, ``Rate--distortion theory for {G}aussian multiterminal source coding systems with several side informations at the decoder,'' \emph{{IEEE} Trans. Inf. Theory}, vol.~51, no.~7, pp. 2577--2593, Jul. 2005.

\bibitem{ElGamal_Kim_2011}
A.~El~Gamal and Y.-H. Kim, \emph{Network Information Theory}.\hskip 1em plus 0.5em minus 0.4em\relax Cambridge University Press, 2011, chapter 11.

\bibitem{Wyner-Ziv1976}
A.~Wyner and J.~Ziv, ``The rate-distortion function for source coding with side information at the decoder,'' \emph{IEEE Transactions on Information Theory}, vol.~22, no.~1, pp. 1--10, 1976.

\bibitem{el2011network}
A.~El~Gamal and Y.-H. Kim, \emph{Network information theory}.\hskip 1em plus 0.5em minus 0.4em\relax Cambridge university press, 2011.

\bibitem{DISCUS1999}
S.~Pradhan and K.~Ramchandran, ``Distributed source coding using syndromes (discus): design and construction,'' in \emph{Proceedings DCC'99 Data Compression Conference (Cat. No. PR00096)}, 1999, pp. 158--167.

\bibitem{Yang2003}
Y.~Yang, S.~Cheng, Z.~Xiong, and W.~Zhao, ``Wyner-ziv coding based on tcq and ldpc codes,'' in \emph{The Thirty-Seventh Asilomar Conference on Signals, Systems \& Computers, 2003}, vol.~1, 2003, pp. 825--829 Vol.1.

\bibitem{Verdu1998}
S.~Verdu, ``Fifty years of shannon theory,'' \emph{IEEE Transactions on Information Theory}, vol.~44, no.~6, pp. 2057--2078, 1998.

\bibitem{van2017neural}
A.~van~den Oord, O.~Vinyals, and K.~Kavukcuoglu, ``Neural discrete representation learning,'' in \emph{Proceedings of the 31st International Conference on Neural Information Processing Systems}, 2017, pp. 6309--6318.

\bibitem{wang2003multiscale}
Z.~Wang, E.~P. Simoncelli, and A.~C. Bovik, ``Multiscale structural similarity for image quality assessment,'' in \emph{The Thrity-Seventh Asilomar Conference on Signals, Systems \& Computers, 2003}, vol.~2.\hskip 1em plus 0.5em minus 0.4em\relax Ieee, 2003, pp. 1398--1402.

\bibitem{kingma2014auto}
\BIBentryALTinterwordspacing
D.~P. Kingma and M.~Welling, ``Auto-encoding variational bayes,'' in \emph{2nd International Conference on Learning Representations, {ICLR} 2014, Banff, AB, Canada, April 14-16, 2014, Conference Track Proceedings}, Y.~Bengio and Y.~LeCun, Eds., 2014. [Online]. Available: \url{http://arxiv.org/abs/1312.6114}
\BIBentrySTDinterwordspacing

\bibitem{kingma2019variational}
\BIBentryALTinterwordspacing
------, ``An introduction to variational autoencoders,'' \emph{Foundations and Trends® in Machine Learning}, vol.~12, no.~4, pp. 307--392, 2019. [Online]. Available: \url{http://dx.doi.org/10.1561/2200000056}
\BIBentrySTDinterwordspacing

\bibitem{townsend2018practical}
\BIBentryALTinterwordspacing
J.~Townsend, T.~Bird, and D.~Barber, ``Practical lossless compression with latent variables using bits back coding,'' in \emph{International Conference on Learning Representations}, 2019. [Online]. Available: \url{https://openreview.net/forum?id=ryE98iR5tm}
\BIBentrySTDinterwordspacing

\bibitem{balle2017endtoend}
J.~Ballé, V.~Laparra, and E.~P. Simoncelli, ``End-to-end optimized image compression,'' 2017.

\bibitem{balle2018variational}
\BIBentryALTinterwordspacing
J.~Ball{\'{e}}, D.~Minnen, S.~Singh, S.~J. Hwang, and N.~Johnston, ``Variational image compression with a scale hyperprior,'' in \emph{6th International Conference on Learning Representations, {ICLR} 2018, Vancouver, BC, Canada, April 30 - May 3, 2018, Conference Track Proceedings}.\hskip 1em plus 0.5em minus 0.4em\relax OpenReview.net, 2018. [Online]. Available: \url{https://openreview.net/forum?id=rkcQFMZRb}
\BIBentrySTDinterwordspacing

\bibitem{NDIC}
N.~Mital, E.~Ozyilkan, A.~Garjani, and D.~Gunduz, ``Neural distributed image compression using common information,'' 2021.

\bibitem{NDIC-CAM}
N.~Mital, E.~\"Ozyilkan, A.~Garjani, and D.~G\"und\"uz, ``Neural distributed image compression with cross-attention feature alignment,'' in \emph{Proceedings of the IEEE/CVF Winter Conference on Applications of Computer Vision (WACV)}, January 2023, pp. 2498--2507.

\bibitem{rezende2014stochastic}
D.~J. Rezende, S.~Mohamed, and D.~Wierstra, ``Stochastic backpropagation and approximate inference in deep generative models,'' in \emph{International conference on machine learning}.\hskip 1em plus 0.5em minus 0.4em\relax PMLR, 2014, pp. 1278--1286.

\bibitem{garbacea2019low}
\BIBentryALTinterwordspacing
C.~Garbacea, A.~v. den Oord, Y.~Li, F.~S.~C. Lim, A.~Luebs, O.~Vinyals, and T.~C. Walters, ``Low bit-rate speech coding with vq-vae and a wavenet decoder,'' \emph{ICASSP 2019 - 2019 IEEE International Conference on Acoustics, Speech and Signal Processing (ICASSP)}, May 2019. [Online]. Available: \url{http://dx.doi.org/10.1109/ICASSP.2019.8683277}
\BIBentrySTDinterwordspacing

\bibitem{razavi2019generating}
\BIBentryALTinterwordspacing
A.~Razavi, A.~van~den Oord, and O.~Vinyals, ``Generating diverse high-fidelity images with vq-vae-2,'' in \emph{Advances in Neural Information Processing Systems}, H.~Wallach, H.~Larochelle, A.~Beygelzimer, F.~d\textquotesingle Alch\'{e}-Buc, E.~Fox, and R.~Garnett, Eds., vol.~32.\hskip 1em plus 0.5em minus 0.4em\relax Curran Associates, Inc., 2019. [Online]. Available: \url{https://proceedings.neurips.cc/paper/2019/file/5f8e2fa1718d1bbcadf1cd9c7a54fb8c-Paper.pdf}
\BIBentrySTDinterwordspacing

\bibitem{roberts2018hierarchical}
A.~Roberts, J.~Engel, C.~Raffel, C.~Hawthorne, and D.~Eck, ``A hierarchical latent vector model for learning long-term structure in music,'' in \emph{International Conference on Machine Learning}.\hskip 1em plus 0.5em minus 0.4em\relax PMLR, 2018, pp. 4364--4373.

\bibitem{ramesh2021zero}
A.~Ramesh, M.~Pavlov, G.~Goh, S.~Gray, C.~Voss, A.~Radford, M.~Chen, and I.~Sutskever, ``Zero-shot text-to-image generation,'' \emph{arXiv preprint arXiv:2102.12092}, 2021.

\bibitem{yu2022scaling}
J.~Yu, Y.~Xu, J.~Y. Koh, T.~Luong, G.~Baid, Z.~Wang, V.~Vasudevan, A.~Ku, Y.~Yang, B.~K. Ayan \emph{et~al.}, ``Scaling autoregressive models for content-rich text-to-image generation,'' \emph{arXiv preprint arXiv:2206.10789}, 2022.

\bibitem{oord2016conditional}
A.~v.~d. Oord, N.~Kalchbrenner, O.~Vinyals, L.~Espeholt, A.~Graves, and K.~Kavukcuoglu, ``Conditional image generation with pixelcnn decoders,'' in \emph{Proceedings of the 30th International Conference on Neural Information Processing Systems}, ser. NIPS'16.\hskip 1em plus 0.5em minus 0.4em\relax Red Hook, NY, USA: Curran Associates Inc., 2016, p. 4797–4805.

\bibitem{vaswani2017}
A.~Vaswani, N.~Shazeer, N.~Parmar, J.~Uszkoreit, L.~Jones, A.~N. Gomez, L.~Kaiser, and I.~Polosukhin, ``Attention is all you need,'' in \emph{Proceedings of the 31st International Conference on Neural Information Processing Systems}, ser. NIPS'17.\hskip 1em plus 0.5em minus 0.4em\relax Red Hook, NY, USA: Curran Associates Inc., 2017, p. 6000–6010.

\bibitem{langdon1984introduction}
G.~G. Langdon, ``An introduction to arithmetic coding,'' \emph{IBM Journal of Research and Development}, vol.~28, no.~2, pp. 135--149, 1984.

\bibitem{ayzikA2020dsin}
S.~Ayzik and S.~Avidan, ``Deep image compression using decoder side information,'' in \emph{Computer Vision - {ECCV} 2020 - 16th European Conference, Glasgow, UK, August 23-28, 2020, Proceedings, Part {XVII}}, vol. 12362, 2020, pp. 699--714.

\bibitem{zhang2023ldmic}
\BIBentryALTinterwordspacing
X.~Zhang, J.~Shao, and J.~Zhang, ``{LDMIC}: Learning-based distributed multi-view image coding,'' in \emph{The Eleventh International Conference on Learning Representations}, 2023. [Online]. Available: \url{https://openreview.net/forum?id=ILQVw4cA5F9}
\BIBentrySTDinterwordspacing

\bibitem{ozyilkan2023neural}
E.~Ozyilkan, J.~Ballé, and E.~Erkip, ``Neural distributed compressor discovers binning,'' 2023.

\bibitem{minnen2018joint}
D.~Minnen, J.~Ball{\'e}, and G.~Toderici, ``Joint autoregressive and hierarchical priors for learned image compression,'' in \emph{Proceedings of the 32nd International Conference on Neural Information Processing Systems}, 2018, pp. 10\,794--10\,803.

\bibitem{theis2017lossy}
L.~Theis, W.~Shi, A.~Cunningham, and F.~Huszár, ``Lossy image compression with compressive autoencoders,'' 2017.

\bibitem{balle2021nonlinear}
\BIBentryALTinterwordspacing
J.~Ballé, P.~A. Chou, D.~Minnen, S.~Singh, N.~Johnston, E.~Agustsson, S.~J. Hwang, and G.~Toderici, ``Nonlinear transform coding,'' \emph{{IEEE} Trans. on Special Topics in Signal Processing}, vol.~15, 2021. [Online]. Available: \url{https://arxiv.org/pdf/2007.03034}
\BIBentrySTDinterwordspacing

\bibitem{chen2018snail}
\BIBentryALTinterwordspacing
X.~Chen, N.~Mishra, M.~Rohaninejad, and P.~Abbeel, ``{P}ixel{SNAIL}: An improved autoregressive generative model,'' in \emph{Proceedings of the 35th International Conference on Machine Learning}, ser. Proceedings of Machine Learning Research, J.~Dy and A.~Krause, Eds., vol.~80.\hskip 1em plus 0.5em minus 0.4em\relax PMLR, 10--15 Jul 2018, pp. 864--872. [Online]. Available: \url{https://proceedings.mlr.press/v80/chen18h.html}
\BIBentrySTDinterwordspacing

\bibitem{geiger2012we}
A.~Geiger, P.~Lenz, and R.~Urtasun, ``Are we ready for autonomous driving? the kitti vision benchmark suite,'' in \emph{2012 IEEE conference on computer vision and pattern recognition}.\hskip 1em plus 0.5em minus 0.4em\relax IEEE, 2012, pp. 3354--3361.

\bibitem{menze2015joint}
M.~Menze, C.~Heipke, and A.~Geiger, ``Joint 3d estimation of vehicles and scene flow,'' \emph{ISPRS annals of the photogrammetry, remote sensing and spatial information sciences}, vol.~2, p. 427, 2015.

\bibitem{ndajah2010ssim}
P.~Ndajah, H.~Kikuchi, M.~Yukawa, H.~Watanabe, and S.~Muramatsu, ``Ssim image quality metric for denoised images,'' in \emph{Proceedings of the 3rd WSEAS International Conference on Visualization, Imaging and Simulation}, ser. VIS '10.\hskip 1em plus 0.5em minus 0.4em\relax Stevens Point, Wisconsin, USA: World Scientific and Engineering Academy and Society (WSEAS), 2010, p. 53–57.

\bibitem{liu2015deep}
Z.~Liu, P.~Luo, X.~Wang, and X.~Tang, ``Deep learning face attributes in the wild,'' in \emph{Proceedings of the IEEE international conference on computer vision}, 2015, pp. 3730--3738.

\bibitem{kingma2018glow}
D.~P. Kingma and P.~Dhariwal, ``Glow: {Generative} flow with invertible 1x1 convolutions,'' in \emph{Neural Information Processing Systems}, 2018, pp. 10\,215--10\,224.

\bibitem{lecun2010mnist}
Y.~LeCun, C.~Cortes, and C.~Burges, ``Mnist handwritten digit database,'' \emph{ATT Labs [Online]. Available: http://yann.lecun.com/exdb/mnist}, vol.~2, 2010.

\bibitem{wang2018atomo}
H.~Wang, S.~Sievert, S.~Liu, Z.~Charles, D.~Papailiopoulos, and S.~Wright, ``Atomo: Communication-efficient learning via atomic sparsification,'' in \emph{Advances in Neural Information Processing Systems}, 2018, pp. 9850--9861.

\bibitem{koloskova2019decentralizeda}
A.~Koloskova, T.~Lin, S.~U. Stich, and M.~Jaggi, ``Decentralized deep learning with arbitrary communication compression,'' \emph{arXiv preprint arXiv:1907.09356}, 2019.

\bibitem{shi2019understanding}
S.~Shi, X.~Chu, K.~C. Cheung, and S.~See, ``Understanding top-k sparsification in distributed deep learning,'' \emph{arXiv preprint arXiv:1911.08772}, 2019.

\bibitem{stich2018sparsified}
S.~U. Stich, J.-B. Cordonnier, and M.~Jaggi, ``Sparsified {SGD} with memory,'' in \emph{Advances in Neural Information Processing Systems}, 2018, pp. 4447--4458.

\bibitem{alistarh2017qsgd}
D.~Alistarh, D.~Grubic, J.~Li, R.~Tomioka, and M.~Vojnovic, ``{QSGD}: {C}ommunication-efficient {SGD} via gradient quantization and encoding,'' in \emph{Advances in Neural Information Processing Systems}, 2017, pp. 1709--1720.

\bibitem{basu2019qsparse}
D.~Basu, D.~Data, C.~Karakus, and S.~Diggavi, ``Qsparse-local-{SGD}: {D}istributed {SGD} with quantization, sparsification and local computations,'' in \emph{Advances in Neural Information Processing Systems}, 2019, pp. 14\,668--14\,679.

\bibitem{shi2019distributed}
S.~Shi, Q.~Wang, K.~Zhao, Z.~Tang, Y.~Wang, X.~Huang, and X.~Chu, ``A distributed synchronous sgd algorithm with global top-k sparsification for low bandwidth networks,'' in \emph{2019 IEEE 39th International Conference on Distributed Computing Systems (ICDCS)}.\hskip 1em plus 0.5em minus 0.4em\relax IEEE, 2019, pp. 2238--2247.

\bibitem{zhang2018unreasonable}
R.~Zhang, P.~Isola, A.~A. Efros, E.~Shechtman, and O.~Wang, ``The unreasonable effectiveness of deep features as a perceptual metric,'' in \emph{Proceedings of the IEEE conference on computer vision and pattern recognition}, 2018, pp. 586--595.

\bibitem{srivastava2017veegan}
A.~Srivastava, L.~Valkov, C.~Russell, M.~U. Gutmann, and C.~Sutton, ``Veegan: Reducing mode collapse in gans using implicit variational learning,'' in \emph{Proceedings of the 31st International Conference on Neural Information Processing Systems}, 2017, pp. 3310--3320.

\bibitem{metz2016unrolled}
L.~Metz, B.~Poole, D.~Pfau, and J.~Sohl-Dickstein, ``Unrolled generative adversarial networks,'' \emph{arXiv preprint arXiv:1611.02163}, 2016.

\bibitem{standard:ldpc}
IEEEStandards, ``{IEEE} draft standard for information technology -- telecommunications and information exchange between systems local and metropolitan area networks -- specific requirements - part 11: Wireless {LAN} medium access control ({MAC}) and physical layer ({PHY}) specifications,'' \emph{IEEE P802.11-REVmd/D1.0, February 2018}, pp. 1--4226, 2018.

\bibitem{matlab:ldpc}
MATLAB, ``{LDPC} encoding and {LDPC} {Q}uasi{C}yclic{M}atrix,'' \url{https://www.mathworks.com/help/comm/ref/ldpcencode.html}, 2021, accessed: 01-25-2023.

\bibitem{TurboAE}
\BIBentryALTinterwordspacing
Y.~Jiang, H.~Kim, H.~Asnani, S.~Kannan, S.~Oh, and P.~Viswanath, ``Turbo autoencoder: Deep learning based channel codes for point-to-point communication channels,'' in \emph{Advances in Neural Information Processing Systems}, H.~Wallach, H.~Larochelle, A.~Beygelzimer, F.~d\textquotesingle Alch\'{e}-Buc, E.~Fox, and R.~Garnett, Eds., vol.~32.\hskip 1em plus 0.5em minus 0.4em\relax Curran Associates, Inc., 2019. [Online]. Available: \url{https://proceedings.neurips.cc/paper_files/paper/2019/file/228499b55310264a8ea0e27b6e7c6ab6-Paper.pdf}
\BIBentrySTDinterwordspacing

\bibitem{KOCodes}
\BIBentryALTinterwordspacing
A.~V. Makkuva, X.~Liu, M.~V. Jamali, H.~Mahdavifar, S.~Oh, and P.~Viswanath, ``Ko codes: inventing nonlinear encoding and decoding for reliable wireless communication via deep-learning,'' in \emph{Proceedings of the 38th International Conference on Machine Learning}, ser. Proceedings of Machine Learning Research, M.~Meila and T.~Zhang, Eds., vol. 139.\hskip 1em plus 0.5em minus 0.4em\relax PMLR, 18--24 Jul 2021, pp. 7368--7378. [Online]. Available: \url{https://proceedings.mlr.press/v139/makkuva21a.html}
\BIBentrySTDinterwordspacing

\bibitem{ProductAE}
\BIBentryALTinterwordspacing
M.~V. Jamali, H.~Saber, H.~Hatami, and J.~H. Bae, ``Productae: Toward training larger channel codes based on neural product codes,'' \emph{ICC 2022 - IEEE International Conference on Communications}, pp. 3898--3903, 2021. [Online]. Available: \url{https://api.semanticscholar.org/CorpusID:238582716}
\BIBentrySTDinterwordspacing

\end{thebibliography}
